\documentclass[sn-mathphys,Numbered,iicol]{sn-jnl}

\usepackage{graphicx}%
\usepackage{multirow}%
\usepackage{amsmath,amssymb,amsfonts}%
\usepackage{amsthm}%
\usepackage{mathrsfs}%
\usepackage[title]{appendix}%
\usepackage{xcolor}%
\usepackage{textcomp}%
\usepackage{manyfoot}%
\usepackage{booktabs}%
\usepackage{algorithm}%
\usepackage{algorithmicx}%
\usepackage{algpseudocode}%
\usepackage{listings}%
\usepackage{fancyhdr}
\usepackage{csquotes}

\usepackage[utf8]{inputenc}  
\usepackage[T1]{fontenc}     
\usepackage{lmodern}         
\usepackage{microtype}       
\usepackage[scaled]{helvet}  
\usepackage[english]{babel}  
\usepackage{graphicx}        
\usepackage{xspace}

\theoremstyle{thmstyleone}%
\theoremstyle{thmstyletwo}%

\theoremstyle{thmstylethree}%

\begin{document}

%

\newcommand{\pp}           {\ensuremath{\mathrm{pp}}\xspace}
\newcommand{\ppbar}        {\mbox{\ensuremath{\mathrm {p\overline{p}}}}\xspace}
\newcommand{\XeXe}         {\ensuremath{\mathrm{\mbox{XeXe}}}\xspace}
\newcommand{\PbPb}         {\ensuremath{\mathrm{\mbox{PbPb}}}\xspace}
\newcommand{\pA}           {\ensuremath{\mathrm{\mbox{pA}}}\xspace}
\newcommand{\pPb}          {\ensuremath{\mathrm{\mbox{pPb}}}\xspace}
\newcommand{\AuAu}         {\ensuremath{\mathrm{\mbox{AuAu}}}\xspace}
\newcommand{\OO}         {\ensuremath{\mathrm{\mbox{OO}}}\xspace}
\renewcommand{\AA}           {\ensuremath{\mathrm{\mbox{AA}}}\xspace}
\newcommand{\dAu}          {\ensuremath{\mathrm{\mbox{dAu}}}\xspace}
\newcommand{\ee}           {\ensuremath{\mathrm{e^{+}e^{-}}}\xspace}
\newcommand{\ep}           {\ensuremath{\mathrm{e^{\pm}p\xspace}}}
\newcommand{\qqbar}        {\ensuremath{\mathrm{q}\overline{\mathrm{q}}}\xspace}

\newcommand{\s}            {\ensuremath{\sqrt{s}}\xspace}
\newcommand{\snn}          {\ensuremath{\sqrt{s_{\mathrm{NN}}}}\xspace}
\newcommand{\pt}           {\ensuremath{p_{\rm T}}\xspace}
\newcommand{\meanpt}       {$\langle p_{\mathrm{T}}\rangle$\xspace}
\newcommand{\ycms}         {\ensuremath{y_{\rm CMS}}\xspace}
\newcommand{\ylab}         {\ensuremath{y_{\rm lab}}\xspace}
\newcommand{\etarange}[1]  {\mbox{$\left | \eta \right |~<~#1$}}
\newcommand{\yrange}[1]    {\mbox{$\left | y \right |~<~#1$}}
\newcommand{\dndy}         {\ensuremath{\mathrm{d}N_\mathrm{ch}/\mathrm{d}y}\xspace}
\newcommand{\dndeta}       {\ensuremath{\mathrm{d}N_\mathrm{ch}/\mathrm{d}\eta}\xspace}
\newcommand{\avdndeta}     {\ensuremath{\langle\dndeta\rangle}\xspace}
\newcommand{\dNdy}         {\ensuremath{\mathrm{d}N_\mathrm{ch}/\mathrm{d}y}\xspace}
\newcommand{\Npart}        {\ensuremath{N_\mathrm{part}}\xspace}
\newcommand{\Ncoll}        {\ensuremath{N_\mathrm{coll}}\xspace}
\newcommand{\dEdx}         {\ensuremath{\textrm{d}E/\textrm{d}x}\xspace}
\newcommand{\RpPb}         {\ensuremath{R_{\rm pPb}}\xspace}
\newcommand{\Tfo}         {\ensuremath{T_\mathrm{fo}}\xspace}
\newcommand{\Th}         {\ensuremath{T_\mathrm{h}}\xspace}
\newcommand{\de}          {\ensuremath{\mathrm{d}}\xspace}
\newcommand{\Raa}          {\ensuremath{R_\mathrm{AA}}\xspace}
\newcommand{\FFc}{\ensuremath{f(\mathrm{c\rightarrow H_{c}})}\xspace}
\newcommand{\zjet}{\ensuremath{z_{\mathrm{||}}^{\mathrm{ch}}}\xspace}
\newcommand{\nineH}        {$\sqrt{s}~=~0.9$~Te\kern-.1emV\xspace}
\newcommand{\seven}        {$\sqrt{s}~=~7$~Te\kern-.1emV\xspace}
\newcommand{\twoH}         {$\sqrt{s}~=~0.2$~Te\kern-.1emV\xspace}
\newcommand{\twosevensix}  {$\sqrt{s}~=~2.76$~Te\kern-.1emV\xspace}
\newcommand{\five}         {$\sqrt{s}~=~5.02$~Te\kern-.1emV\xspace}
\newcommand{\twosevensixnn}{$\sqrt{s_{\mathrm{NN}}}~=~2.76$~Te\kern-.1emV\xspace}
\newcommand{\fivenn}       {$\sqrt{s_{\mathrm{NN}}}~=~5.02$~Te\kern-.1emV\xspace}
\newcommand{\LT}           {L{\'e}vy-Tsallis\xspace}
\newcommand{\GeVc}         {\ensuremath{\mathrm{GeV}/c}\xspace}
\newcommand{\MeVc}         {\ensuremath{\mathrm{MeV}/c}\xspace}
\newcommand{\TeV}          {\ensuremath{\mathrm{TeV}}\xspace}
\newcommand{\GeV}          {\ensuremath{\mathrm{GeV}}\xspace}
\newcommand{\MeV}          {\ensuremath{\mathrm{MeV}}\xspace}
\newcommand{\GeVmass}      {\ensuremath{\mathrm{GeV}/c^2}\xspace}
\newcommand{\MeVmass}      {\ensuremath{\mathrm{MeV}/c^2}\xspace}
\newcommand{\lumi}         {\ensuremath{\mathcal{L}}\xspace}
\newcommand{\mub}         {\ensuremath{\mu\mathrm{b}}\xspace}
\newcommand{\Taa}          {\ensuremath{\langle T_\mathrm{AA}\rangle}\xspace}
\newcommand{\Diffs}        {\ensuremath{D_{s}}\xspace}
\newcommand{\vtwo}          {\ensuremath{v_\mathrm{2}}\xspace}
\newcommand{\ITS}          {\rm{ITS}\xspace}
\newcommand{\TOF}          {\rm{TOF}\xspace}
\newcommand{\ZDC}          {\rm{ZDC}\xspace}
\newcommand{\ZDCs}         {\rm{ZDCs}\xspace}
\newcommand{\ZNA}          {\rm{ZNA}\xspace}
\newcommand{\ZNC}          {\rm{ZNC}\xspace}
\newcommand{\SPD}          {\rm{SPD}\xspace}
\newcommand{\SDD}          {\rm{SDD}\xspace}
\newcommand{\SSD}          {\rm{SSD}\xspace}
\newcommand{\TPC}          {\rm{TPC}\xspace}
\newcommand{\TRD}          {\rm{TRD}\xspace}
\newcommand{\VZERO}        {\rm{V0}\xspace}
\newcommand{\VZEROA}       {\rm{V0A}\xspace}
\newcommand{\VZEROC}       {\rm{V0C}\xspace}
\newcommand{\Vdecay} 	   {\ensuremath{V^{0}}\xspace}

\newcommand{\pip}          {\ensuremath{\pi^{+}}\xspace}
\newcommand{\pim}          {\ensuremath{\pi^{-}}\xspace}
\newcommand{\kap}          {\ensuremath{\rm{K}^{+}}\xspace}
\newcommand{\kam}          {\ensuremath{\rm{K}^{-}}\xspace}
\newcommand{\pbar}         {\ensuremath{\rm\overline{p}}\xspace}
\newcommand{\kzero}        {\ensuremath{{\rm K}^{0}_{\rm{S}}}\xspace}
\newcommand{\lmb}          {\ensuremath{\Lambda}\xspace}
\newcommand{\almb}         {\ensuremath{\overline{\Lambda}}\xspace}
\newcommand{\Om}           {\ensuremath{\Omega^-}\xspace}
\newcommand{\Mo}           {\ensuremath{\overline{\Omega}^+}\xspace}
\newcommand{\X}            {\ensuremath{\Xi^-}\xspace}
\newcommand{\Ix}           {\ensuremath{\overline{\Xi}^+}\xspace}
\newcommand{\Xis}          {\ensuremath{\Xi^{\pm}}\xspace}
\newcommand{\Oms}          {\ensuremath{\Omega^{\pm}}\xspace}
\newcommand{\degree}       {\ensuremath{^{\rm o}}\xspace}
\newcommand{\Hc}           {\ensuremath{\mathrm{H_c}}\xspace}
\newcommand{\Dzero}        {\ensuremath{\mathrm{D^0}}\xspace}
\newcommand{\Dplus}        {\ensuremath{\mathrm{D^+}}\xspace}
\newcommand{\Dminus}        {\ensuremath{\mathrm{D^-}}\xspace}

\newcommand{\Dstar}        {\ensuremath{\mathrm{D^{*+}}}\xspace}
\newcommand{\Dstarzero}        {\ensuremath{\mathrm{D^{*0}}}\xspace}
\newcommand{\Ds}           {\ensuremath{\mathrm{D_s^+}}\xspace}
 \newcommand{\Bs}           {\ensuremath{\mathrm{B_s^0}}\xspace}
 \newcommand{\Bplus}           {\ensuremath{\mathrm{B^+}}\xspace}
  \newcommand{\Bzero}           {\ensuremath{\mathrm{B^0}}\xspace}
  \newcommand{\Bc}           {\ensuremath{\mathrm{B^+_c}}\xspace}

\newcommand{\Dsstar}{\ensuremath{\mathrm{D_s^{*+}}}\xspace}

\newcommand{\Dsminus}    
{\ensuremath{\mathrm{D_s^-}}\xspace}
\newcommand{\Lc}           {\ensuremath{\Lambda_\mathrm{c}^+}\xspace}
\newcommand{\Lcgeneric}           {\ensuremath{\Lambda_\mathrm{c}}\xspace}
\newcommand{\Lcminus}           {\ensuremath{\Lambda_\mathrm{c}^-}\xspace}
\newcommand{\LcExcitedOne}{\ensuremath{\Lambda_\mathrm{c}(2595)^+}\xspace}
\newcommand{\LcExcitedTwo}{\ensuremath{\Lambda_\mathrm{c}(2625)^+}\xspace}
\newcommand{\SigmacZero}           {\ensuremath{\Sigma_\mathrm{c}(2455)^{0}}\xspace}
\newcommand{\SigmacZeroPlus}           {\ensuremath{\Sigma_\mathrm{c}(2455)^{0,++}}\xspace}
\newcommand{\SigmacZeroExcited}           {\ensuremath{\Sigma_\mathrm{c}(2520)^{0}}\xspace}

\newcommand{\Sigmac}           {\ensuremath{\Sigma_\mathrm{c}^{0,+,++}}\xspace}
\newcommand{\Sigmacgeneric}           {\ensuremath{\Sigma_\mathrm{c}}\xspace}

\newcommand{\XicZero}      {\ensuremath{\Xi_\mathrm{c}^0}\xspace}
\newcommand{\XicPlus}      {\ensuremath{\Xi_\mathrm{c}^+}\xspace}
\newcommand{\XicPlusZero}  {\ensuremath{\Xi_\mathrm{c}^{0,+}}\xspace}
\newcommand{\Omegac}       {\ensuremath{\Omega_\mathrm{c}^0}\xspace}
\newcommand{\Jpsi}         {\ensuremath{\mathrm{J}/\psi}\xspace}
\newcommand{\PsiTwos}         {\ensuremath{\psi(\mathrm{2S})}\xspace}
\newcommand{\UpsilonOneS}         {\ensuremath{\mathrm{\Upsilon (1S)}}\xspace}
\newcommand{\ccbar}        {\ensuremath{\mathrm{c\overline{c}}}\xspace}
\newcommand{\bbbar}        {\ensuremath{\mathrm{b\overline{b}}}\xspace}

\newcommand{\Xib}          {\ensuremath{\Xi_\mathrm{b}^{0,-}}\xspace}
\newcommand{\DzerotoKpi}   {\ensuremath{\mathrm{D^0\to K^-\pi^+}}}
\newcommand{\Lb}     {\ensuremath{\Lambda_\mathrm{b}^0}\xspace}
\newcommand{\Lbbar}           {\ensuremath{\bar{\Lambda}_\mathrm{b}^0}\xspace}

\newcommand{\lowptbin}{\ensuremath{0<\pt<1}~\GeVc}

\newcommand{\LcD} {\ensuremath{\Lc/\Dzero}\xspace}
\newcommand{\QQbar}           {\ensuremath{\mathrm{Q\bar{Q}}\xspace}}

\newcommand{\tamu}         {\textsc{tamu}\xspace}
\newcommand{\pythiasix}    {\textsc{pythia6}\xspace}
\newcommand{\herwig}    {\textsc{herwig}\xspace}
\newcommand{\pythiaeight}  {\textsc{pythia8}\xspace}
\newcommand{\pythiaeightprecise}{\textsc{pythia8.243}\xspace}
\newcommand{\pythiasixprecise}{\textsc{pythia6.4.25}\xspace}
\newcommand{\pythia}       {\textsc{pythia}\xspace}
\newcommand{\hijing}       {\textsc{hijing}\xspace}
\newcommand{\hijingprecise}{\textsc{hijing v1.383}\xspace}
\newcommand{\fonll}        {\textsc{fonll}\xspace}
\newcommand{\evtgen}       {\textsc{EvtGen}\xspace}

\title[Article Title]{QCD challenges from pp to AA collisions - 4th edition}

\author[1]{Javira Altmann}
\author[2]{Carlota Andres}
\author[3]{Anton Andronic}
\author[4]{Federico Antinori}
\author[5]{Pietro Antonioli}
\author[6]{Andrea Beraudo}
\author[7]{Eugenio Berti}
\author[6,8]{Livio Bianchi}
\author[9]{Thomas Boettcher}
\author[10]{Lorenzo Capriotti}
\author[11]{Peter Christiansen}
\author[12]{Jesus Guillermo Contreras Nu\~no}
\author[13]{Leticia Cunqueiro Mendez}
\author[14]{Cesar da Silva}
\author[4]{Andrea Dainese}
\author[15]{Hans Peter Dembinski}
\author[16,17]{David Dobrigkeit Chinellato}
\author[18]{Andrea Dubla}
\author[19,20]{Mattia Faggin}
\author[21]{Chris Flett}
\author[22,23]{Vincenzo Greco}
\author[24]{Ilia Grishmanovskii}
\author[2]{Jack Holguin}
\author[25]{Yuuka Kanakubo}
\author[25]{Dong Jo Kim}
\author[26]{Ramona Lea}
\author[27]{Su Houng Lee}
\author[28]{Saverio Mariani}
\author[29]{Adam Matyja}
\author[30]{Aleksas Mazeliauskas}
\author[22,31]{Vincenzo Minissale}
\author[28]{Andreas Morsch}
\author[22,31]{Lucia Oliva}
\author[6,8]{Luca Orusa}
\author[25]{Petja Paakkinen}
\author[6,32,33]{Daniel Pablos}
\author[34]{Guy Pai\'{c}}
\author[35]{Tanguy Pierog}
\author[22,23]{Salvatore Plumari}
\author[6]{Francesco Prino}
\author[4]{Andrea Rossi}
\author[4]{Lorenzo Sestini}
\author[1]{Peter Skands}
\author[36]{Olga Soloveva}
\author[4,19]{Francesca Soramel}
\author[28]{Alba Soto Ontoso}
\author[37]{Martin Spousta}
\author[28]{Andre Govinda Stahl Leiton}
\author[38]{Jiayin Sun}
\author[39]{Adam Takacs}
\author[8,40]{Stefano Trogolo}
\author[4]{Rosario Turrisi}
\author[41]{Marta Verweij}
\author[11]{Vytautas Vislavicius}
\author[42]{Jing Wang}
\author[43]{Klaus Werner}
\author[20]{Valentina Zaccolo}
\author[44]{Mingyu Zhang}
\author[4]{Jianhui Zhu}
\author[4,19]{Davide Zuliani}

\affil[1]{Monash University, Melbourne, Australia}
\affil[2]{CPHT, CNRS, Ecole polytechnique, Institut Polytechnique de Paris, Palaiseau, France}
\affil[3]{Universit\"at M\"unster, Germany}
\affil[4]{INFN, Sezione di Padova, Italy}
\affil[5]{Università and INFN Bologna, Italy}
\affil[6]{INFN, Sezione di Torino, Italy}
\affil[7]{INFN, Sezione di Firenze, Italy}
\affil[8]{Università di Torino, Italy}
\affil[9]{University of Cincinnati, US}
\affil[10]{Università and INFN Ferrara, Italy}
\affil[11]{Lund University, Sweden}
\affil[12]{Czech Technical University in Prague, Czech Republic}
\affil[13]{Università Sapienza, Roma, Italy}
\affil[14]{Los Alamos National Lab, US}
\affil[15]{Technische Universität Dortmund, Germany}
\affil[16]{University of Campinas UNICAMP, Brasil}
\affil[17]{Stefan Meyer Institute of the Austrian Academy of Sciences}
\affil[]{\newline \newline \newline \small Contacts: Andrea Rossi (\href{andrea.rossi@pd.infn.it}{andrea.rossi@pd.infn.it}) and Lorenzo Sestini (\href{lorenzo.sestini@pd.infn.it}{lorenzo.sestini@pd.infn.it})}
\affil[18]{GSI Helmholtzzentrum für Schwerionenforschung, Darmstadt, Germany}
\affil[19]{Università di Padova, Italy}
\affil[20]{Università and INFN Trieste, Italy}
\affil[21]{Université Paris-Saclay, CNRS, IJCLab, Orsay, France}
\affil[22]{Università di Catania, Italy}
\affil[23]{INFN-LNS, Catania, Italy}
\affil[24]{Institut für Theoretische Physik, Frankfurt, Germany}
\affil[25]{University of Jyväskylä and Helsinki Institute of Physics, Finland}
\affil[26]{Universita di Brescia, Italy}
\affil[27]{Yonsei University, Seoul, Korea}
\affil[28]{CERN, Geneve, Switzerland}
\affil[29]{The Henryk Niewodniczanski Institute of Nuclear Physics, Polish Academy of Sciences, Cracow, Poland}
\affil[30]{Ruprecht Karls Universitaet Heidelberg, Germany}
\affil[31]{INFN Sezione di Catania, Italy}
\affil[32]{Departamento de F\'isica, Universidad de Oviedo, Avda, Oviedo, Spain}
\affil[33]{Instituto Universitario de Ciencias y Tecnolog\'ias Espaciales de Asturias (ICTEA), Oviedo, Spain}
\affil[34]{Universidad Nacional Autonoma, Ciudad de México, Mexico}
\affil[35]{Karlsruhe Institute of Technology, Institut für Astroteilchenphysik, Germany}
\affil[36]{Helmholtz Research Academy Hesse for FAIR, Goethe University Frankfurt, Germany}
\affil[37]{Charles University, Prague, Czech Republic}
\affil[38]{INFN Sezione di Cagliari}
\affil[39]{University of Bergen, Norway}
\affil[40]{University of Houston, US}
\affil[41]{Utrecht University, Netherlands}
\affil[42]{Massachusetts Institute of Technology, Cambridge, US}
\affil[43]{SUBATECH, Nantes University - IN2P3/CNRS - IMT Atlantique, Nantes, France}
\affil[44]{Central China Normal University, Wuhan, China}


\abstract{This paper is a write-up of the ideas that were presented, developed and discussed at the fourth International Workshop on QCD Challenges from pp to AA, which took place in February 2023 in Padua, Italy. The goal of the workshop was to focus on some of the open questions in the field of high-energy heavy-ion physics and to stimulate the formulation of concrete suggestions for making progresses on both the experimental and theoretical sides. The paper gives a brief introduction to each topic and then summarizes the primary results.}

\maketitle

\section{Introduction} 
\label{sec:intro}
In hadronic collisions complex many-body systems of strongly-interacting particles are produced. The strong interaction, described in the Standard Model by quantum chromodynamics (QCD), determines the properties of the systems formed in these collisions and the evolution of the observed phenomena with the system complexity. %
In collisions of heavy nuclei (AA) at ultra-relativistic energies a system of deconfined quarks and gluons, the quark-gluon plasma (QGP), is formed. The existence of a QGP state is expected from QCD calculations on the lattice~\cite{Karsch:2006xs,Borsanyi:2010bp,Borsanyi:2013bia,Bazavov:2011nk}. The search of evidences of QGP formation and the study of its properties have shaped the high-energy heavy-ion physics program at accelerators in the last 30 years~\cite{NA50:2000brc, WA97:1999uwz,BRAHMS:2004adc,PHENIX:2004vcz,PHOBOS:2004zne,STAR:2005gfr,Schutz:2011zz,Roland:2014jsa,Braun-Munzinger:2015hba,ALICE:2022wpn} and traced the road for future experiments~\cite{PHENIX:2015siv,NA60:2022sze,AbdulKhalek:2021gbh,ALICE:2022wwr}.
Most of the observables and probes used to investigate the QGP were first studied in detail in more elementary, smaller, systems, like electron--positron ($\mathrm{e^{-}e^{+}}$), proton--proton (\pp), and proton--nucleus (\pA) collisions. These studies were fundamental for the comprehension of many aspects of the strong force and the development of QCD. On top of providing a baseline for interpreting the measurements in AA collisions, pp and pA collisions can be exploited to study the onset of phenomena ascribed to QGP in AA collisions.  
The goal of the workshop series ``QCD challenges from pp to AA collisions'' is to bring together experimental and theoretical physicists involved in the study of strong-interaction phenomena, mainly at high-energy particle colliders, with the aim of identifying the main open questions in the field and proposing new ideas and directions of research for addressing them. The focus is on those phenomena in hadronic collisions that are possibly sensitive to the size of the interacting system.

The $4^{\mathrm{th}}$ edition of the workshop was organized in Padua, Italy, from 13 February to 17 February 2023. It consisted in plenary and round-table sessions related to the following six discussion tracks:
\begin{enumerate}
    \item Initial state and ultra-peripheral collisions
    \item Jet production and properties in pp collisions and in the medium
    \item Event properties and hydro in small and large systems
    \item Hadronization of light and heavy flavour across collision systems
    \item Energy loss and transport in the medium and in small systems
    \item QCD and astrophysics
\end{enumerate}

In plenary sessions invited speakers gave an overview on their specific field of interest. During round-table sessions, one per track, small groups (7-8 persons) discussed in detail new ideas and studies from both the experimental and theoretical sides. Moreover, special round-table sessions where groups from two or more tracks discussed together about possible synergies have been organized. The conveners of each track wrapped up the main ideas emerged and presented the outcome of the discussion in the plenary session on the last day of the workshop. In this document the outcome of these discussions is reported. 

A section is dedicated to each track. The choice of the tracks was determined by considering the latest advances and open questions in the field, briefly outlined in what follows. In recent years, an intensive experimental campaign of measurements of jet and heavy-flavour (charm and beauty) production, azimuthal anisotropy, and correlations was carried out, giving an unprecedented insight into the partonic interactions in the medium and setting important constraints for the understanding of partonic energy loss and of heavy-quark transport. This campaign will continue in the next years, also exploiting detector upgrades, with novel measurements of jet substructure and with new and precise measurements of several charm and beauty hadron species, in particular of baryons. A prerequisite for the interpretation of production measurements will be the characterisation of the nuclear parton distribution functions (nPDF) and of initial-state effects. These topics are discussed in Sections~\ref{sec:initialstateUP},~\ref{sec:jets}, and~\ref{sec:energyloss}.

In the last decade, phenomena like long-range flow-like angular correlations, studied in heavy-ion collisions and typically interpreted as a consequence of the formation and expansion of a medium have been observed, with different magnitude, in pp and pA collisions systems.  A major challenge for the future is understanding the onset of these phenomena across collision systems, their dependence on event properties like particle multiplicity, the role played by multiple parton interactions (MPI), colour reconnection, and fluctuations in the initial state, as well as explaining the absence of evidences for other phenomena like partonic energy loss that naturally accompany flow observations in heavy-ion collisions. This aspect, which was discussed in most sessions, is one of the main subjects of Sections~\ref{sec:hydro}~and~\ref{sec:energyloss}. In recent years, the paradigm that heavy-quark hadronisation should proceed similarly in $\mathrm{e^{-}e^{+}}$ and pp collisions, which motivated the usage of a factorisation approach for calculating charm- and beauty-hadron production cross sections, has been severely questioned by the observation that charm and beauty baryon production relative to that of mesons is significantly larger in pp collisions than in $\mathrm{e^{-}e^{+}}$ collisions. Tracing the modification of the hadronisation process in different hadronic environments, from pp to AA collisions, and the possible emergence of quark coalescence as a relevant hadronisation process already in small collision systems, is a major goal for the experiments. This topic is the subject of Section~\ref{sec:hadronization}.

Many QCD-related measurements performed at hadronic colliders can provide important pieces of information to answer open questions related to cosmic-ray physics and astrophysical objects. The outcome of the discussions emanating from this theme is reported in Section~\ref{sec:astrophysics}.  Section~\ref{sec:summary} closes the document with a summary and an outlook. 


We remark that the choice of topics addressed in the following sections does not aspire to systematically review all open 
questions in the field. It is rather driven by the most important points presented by the participants at the workshop and the results of discussions and brainstorming.  

\section{Initial state and ultraperipheral collisions}
\label{sec:initialstateUP}
The colliding particles in high-energy hadron collisions are dense gluonic
systems. As a result, describing the initial state of high-energy hadron
collisions requires understanding QCD at high gluon densities. These conditions
can be studied by probing the structure of heavy nuclei at low $x$ and $Q^2$.
The longitudinal structure of nuclei can be described using nPDFs~\cite{AbdulKhalek:2022fyi,Eskola:2021nhw,Duwentaster:2022kpv,Helenius:2021tof}. In nPDF analyses, the nucleon-to-proton PDF
ratio is parameterized as a function of $x$, and evolution in $Q^2$ is governed
by the linear DGLAP equation. At low $x$ and $Q^2$, however, the spatial
separation between gluons may be smaller than the size scale of the probing
interaction. In this low-$x$ regime, nonlinearity may affect the evolution of
parton densities, and coherent interactions with multiple gluons could become
important. These so-called saturation effects are described by the color-glass
condensate (CGC) effective field theory\,\cite{McLerran:1993ka}. Experimentally, the low-$x$ regime is
most commonly accessed at hadron colliders using two classes of observables:
forward inclusive particle production in pA collisions and
photoproduction in ultraperipheral collisions (UPCs). Observing and studying
nonlinear QCD effects is one of the primary goals of initial state physics.

\subsection{Forward measurements and low-$x$ suppression}
\begin{figure*}
    \centering
    \includegraphics[width=0.48\textwidth]{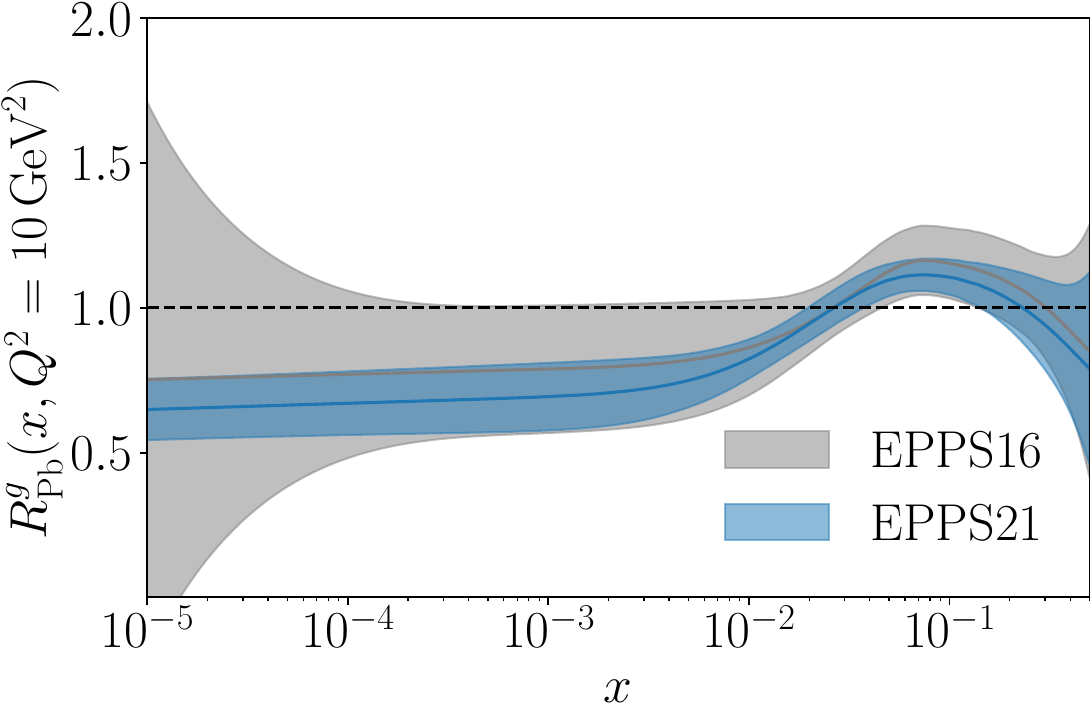}
    \includegraphics[width=0.48\textwidth]{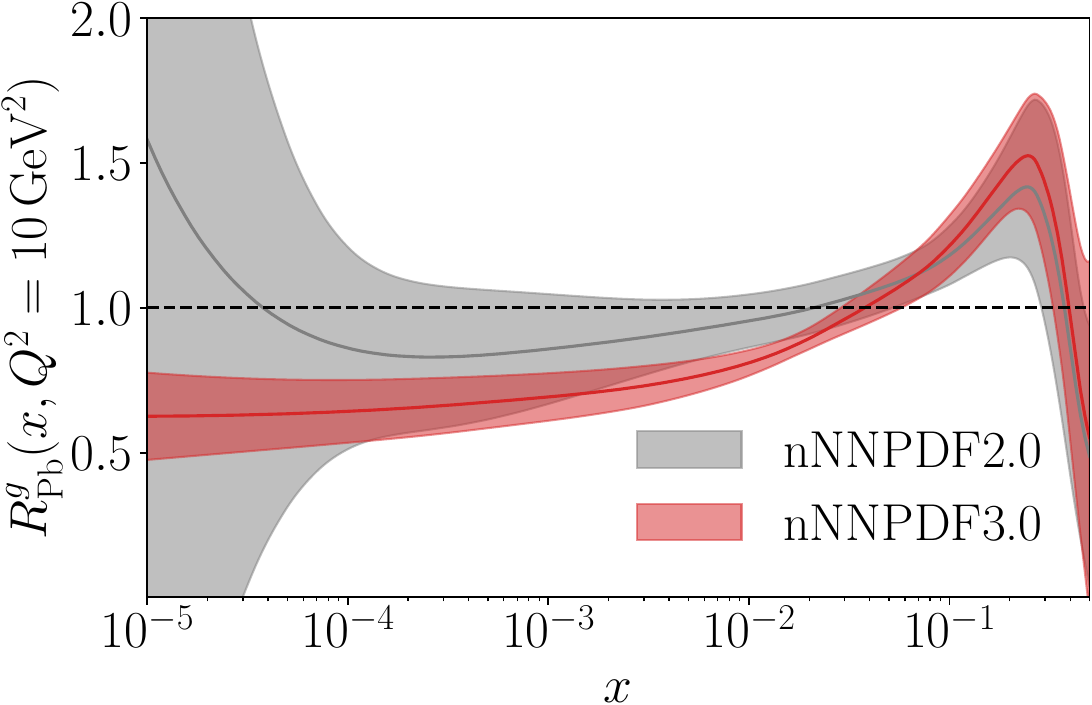}
    \caption{Comparisons of current- and previous-generation gluon nuclear modification factors $R^g_{{\rm pPb}}$ in $^{208}{\rm Pb}$ from the (left) EPPS\,\cite{Eskola:2016oht,Eskola:2021nhw} and (right) nNNPDF\,\cite{AbdulKhalek:2020yuc,AbdulKhalek:2022fyi} collaborations. The shaded regions show the $68\%~{\rm CL}$ uncertainties. In both cases, the fit results shown in gray do not use LHCb $\Dzero$ data\,\cite{LHCb:2017yua}, while the results shown in color do use these data.}
    \label{initialstate:fig1}
\end{figure*}

Forward detectors at high-energy hadron colliders are ideal tools for studying
the low-$x$ regime. The LHCb detector, for example, can probe $x$ below
$10^{-5}$ in charm production in pPb collisions. LHCb measurements of
the $\Dzero$ nuclear modification factor in pPb collisions at $\sqrt{s_{\rm
NN}}=5\,{\rm TeV}$\,\cite{LHCb:2017yua} have been used in state-of-the-art nPDF fits to obtain
precise descriptions of the gluon distribution at low $x$~\cite{Eskola:2021nhw,AbdulKhalek:2022fyi}. The improvement in precision over previous-generation nPDF fits is illustrated in Fig.~\ref{initialstate:fig1}. The LHCb $\Dzero$ data
offer clear evidence of suppression of the low-$x$ gluon density in heavy
nuclei. The resulting suppression is consistent with both collinearly factorizable shadowing of the gluon
nPDF and gluon saturation in the CGC framework. As a result, it is now clear that observing nonlinear
QCD effects in forward particle production will require high experimental
precision across a wide range of observables.

In order to push to lower $x$ and $Q^2$, the LHCb collaboration has studied both
charged hadron\,\cite{LHCb:2021vww} and neutral pion\,\cite{LHCb:2022tjh} production in pPb collisions. The forward
results again agree with both nPDF and CGC predictions. The ALICE collaboration
has also measured the nuclear modification of charged hadrons and $\pi^0$ mesons in
pPb collisions at central rapidity\,\cite{ALICE:2018vuu,ALICE:2018vhm,ALICE:2021est}. The ALICE pPb results show
signs of suppression at $p_{\rm T}\lesssim 2\,{\rm GeV}$, as opposed to the
Cronin-like peak observed in PbPb collisions. The CMS
collaboration has also measured the nuclear modification of charged hadrons\,\cite{CMS:2016xef},
observing a similar suppression to that observed by ALICE at low $p_{\rm T}$.
Altogether, the LHCb, ALICE, and CMS results probe the nuclear gluon density
from $x\lesssim10^{-5}$ to $x\gtrsim{10^{-1}}$ over a wide range of $Q^2$.

In addition to nonlinear effects in the evolution of parton densities, gluon
saturation can also result in multiple scattering with low-$x$ gluons, leading
to modifications of particle correlations at forward rapidity in pPb
collisions. Both the PHENIX and STAR experiments have studied di-pion correlations
at forward rapidity, observing a suppression of back-to-back low-$p_{\rm T}$
di-pion pairs in deuteron-gold (dAu) and pA collisions\,\cite{STAR:2021fgw,PHENIX:2018foa}. The STAR collaboration
also observed that this suppression scales with $A^{1/3}$, consistent with
saturation predictions. The ATLAS collaboration sees a similar suppression of
back-to-back dijets in pPb collisions\,\cite{ATLAS:2019jgo}.

An ideal probe of the nuclear gluon distribution is direct photon production in
pA collisions\,\cite{Jalilian-Marian:2012wwi}. Direct photons probe the gluon nPDF at leading order
and could provide access to the saturation regime at low $p_{\rm T}$ and forward
rapidity. The nuclear modification of low-$p_{\rm T}$ direct photons could
reveal nonlinear QCD effects, and photon-hadron correlations could provide
evidence for gluon saturation. The ALICE collaboration is planning to install a
high-granularity forward electromagnetic calorimeter (FOCAL) that will allow for
measurements of forward direct photon production in Run 4 of the LHC\,\cite{ALICE:2020mso}. The LHCb
collaboration is also planning on installing tracking stations inside of its
dipole magnet before Run 4, dramatically increasing its acceptance for
low-momentum converted photons\,\cite{LHCbCollaboration:2776420}.

\subsection{Probing low $x$ with vector-meson photoproduction in UPCs}
Studies of photoproduction of vector mesons in UPCs have provided complementary
probes of the gluonic structure of nucleons. In UPCs, the electric field of a
nucleus produces quasi-real photons, which can interact coherently with the
colliding nucleus. The photon flux scales with $Z^2$, where $Z$ is the atomic
number of the nucleus. Furthermore, hadronic interactions are suppressed in UPCs
due to the large impact parameter of the collision, which exceeds the sum of the
radii of the colliding nuclei. This results in an experimentally clean signal.
Vector meson photoproduction is sensitive to the the gluon density of the nucleus
at leading order at $x=me^{\pm y}/\sqrt{s_{\rm NN}}$, where $m$ is the vector
meson's mass and $y$ is its rapidity. As a result, vector meson photoproduction
in high-energy UPCs can probe gluon densities at extremely low $x$\,\cite{Contreras:2015dqa,Klein:2019qfb}.

\begin{figure*}
    \centering
    \includegraphics[width=0.49\textwidth]{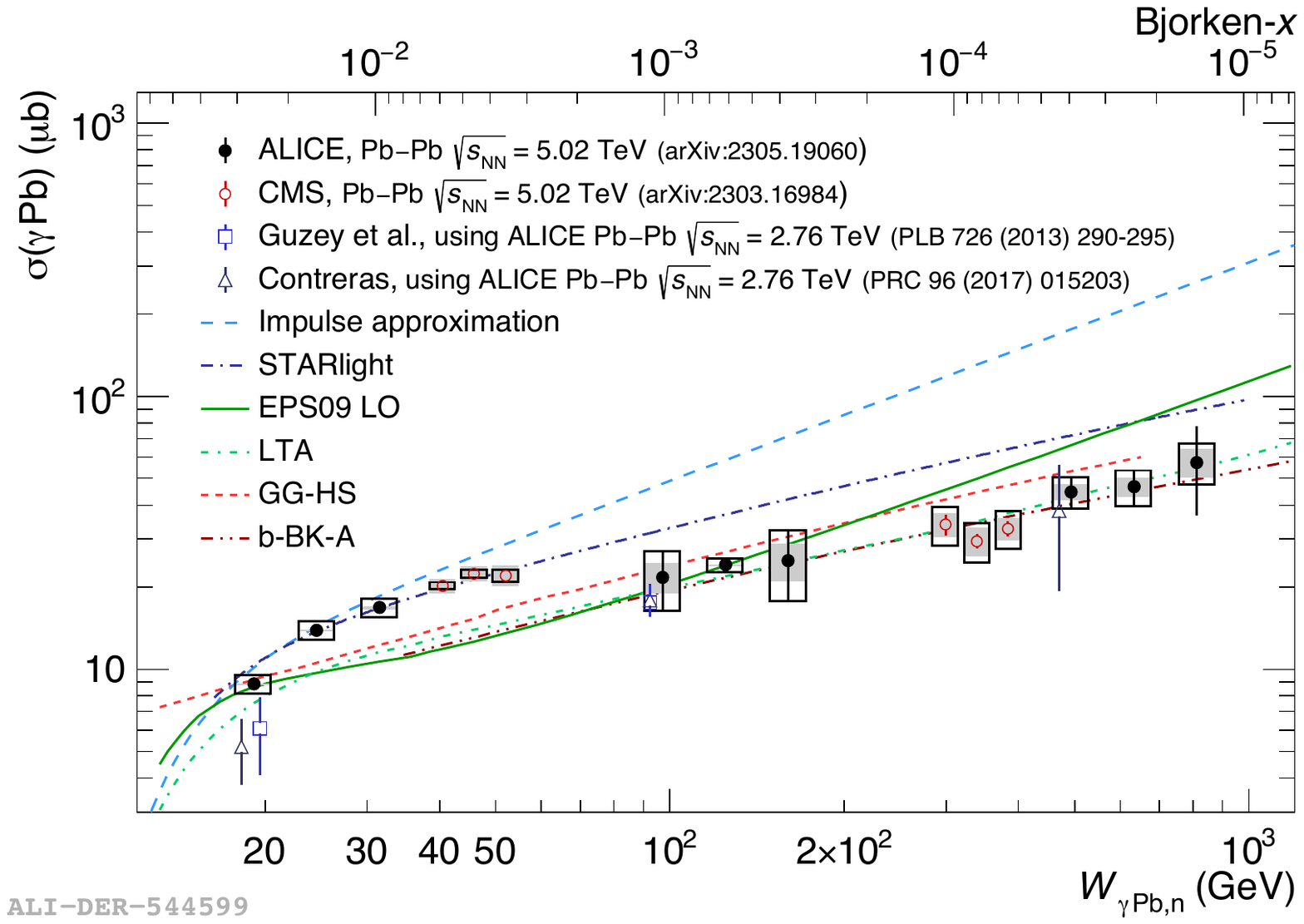}
    \includegraphics[width=0.47\textwidth]{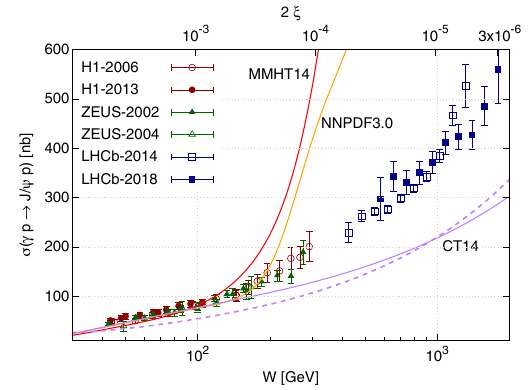}
    \caption{$\Jpsi$ photoproduction cross sections for $\gamma$Pb (left) from ALICE\,\cite{ALICE:2023jgu} and CMS\,\cite{CMS:2023snh} measured using nuclear breakup categories to unambiguously determine the energy dependence and for $\gamma$p (right) from HERA and LHCb measurements versus NLO predictions~\cite{Flett:2019pux}.}
    \label{initialstate:fig2}
\end{figure*}

Coherent vector meson photoproduction has been recently measured in heavy-ion
collisions for
$\rho^0$~\cite{ALICE:2020ugp,ALICE:2021jnv,STAR:2022wfe},
$\Jpsi$~\cite{ALICE:2021gpt,LHCb:2021bfl,LHCb:2022ahs,CMS-PAS-HIN-22-002},
and $\PsiTwos$~\cite{ALICE:2021gpt,LHCb:2022ahs}. The results cover a
wide range in rapidity and show partial agreement with theoretical models. 
The heavy-ion cross section can be used to calculate photonuclear cross section. However, this
calculation suffers from an ambiguity because the photon could be emitted by
either nucleus, resulting in two possible $x$ values for each data point. Two
solutions have been proposed to resolve this ambiguity. First, vector meson
photoproduction can be measured in multiple nuclear breakup classes, which can
then be used to extract photoproduction cross sections at both possible values
of $x$~\cite{Guzey:2013jaa}. This method has been used by both the ALICE and CMS
collaborations to extract low-$x$ photoproduction cross sections\,\cite{ALICE:2023jgu,CMS:2023snh}, which are shown in Fig.~\ref{initialstate:fig2} (left). Second,
coherent production in peripheral and ultraperipheral collisions can be
combined~\cite{Contreras:2016pkc}. Coherent production in peripheral heavy-ion
collisions has been measured by both the ALICE and LHCb collaborations. Both
methods allow for access of unique $x$ range down to $x\sim 10^{-5} - 10^{-6}$.
The experimental results show strong gluon suppression relative to the Impulse
Approximation~\cite{Klein:2016yzr}, which can be best described by
models with either shadowing or
saturation effects~\cite{Accardi:2012qut}.  Additionally, the
$|t|$-dependence of the photonuclear cross section, where $|t|\simeq p_{\rm T}^2$,
is directly sensitive to the spatial distribution of gluons in the
nucleus~\cite{Cepila:2016uku,ALICE:2023mfc,ALICE:2021tyx}.

The gluon distribution in the proton can be studied using vector meson
photoproduction in pp or pPb collisions. The results of
LHCb~\cite{LHCb:2014acg,LHCb:2018rcm} and
ALICE~\cite{ALICE:2014eof,ALICE:2018oyo,Herman:2023mzg} extend the exclusive $\Jpsi$
Bjorken-$x$ photonuclear cross section coverage down to $\sim 10^{-6}$. However,
no clear indication of gluon saturation at low $x$ is visible, and data
follow trends observed in $ep$ collisions at HERA~\cite{H1:2005dtp,H1:2013okq,ZEUS:2002wfj}. A similar
trend is visible in $Y(nS)$~\cite{LHCb:2015wlx,CMS:2018bbk} and
$\rho^0$~\cite{CMS:2019awk} meson data from the CMS collaboration. Deviation
from this trend might be visible in the dissociative vector meson
photoproduction cross section at higher collision energies reachable during Runs
3 and 4 of the LHC.

\subsection{Progress in higher-order calculations for $\Jpsi$ photoproduction}
The theoretical interpretation of vector meson photoproduction data from UPCs is
limited by large factorization scale uncertainties~\cite{Ivanov:2004vd,Eskola:2022vaf,Eskola:2022vpi,Eskola:2023oos}. Furthermore, the LO
and NLO results have opposite signs. This contrasts with the expectations of a
forward $t=0$ elastic scattering amplitude based on Regge theory, and in general
hints at poor perturbative stability of the amplitude. 
This, together with the sensitivity of the process to generalized parton
distributions, poses a major obstacle to including exclusive heavy vector meson
photoproduction data in global PDF analyses.

Recent theoretical work has aimed to overcome these challenges. 
PDFs can be related to GPDs using the so-called Shuvaev integral transform at moderate-to-low values of the skewness parameter $\xi\sim x/2$, which quantifies the fraction of longitudinal momentum transfer from the initial state hadron to that in the final state~\cite{Shuvaev:1999ce,Shuvaev:1999fm,Golec-Biernat:1999trj}.
The conventional collinear factorization coefficient functions
for exclusive heavy vector meson photoproduction at NLO can then be supplemented with
corrections of $\mathcal{O}(Q_0^2/\mu^2)$, where $Q_0$ is the GPD input scale
and $\mu$ the factorization scale~\cite{Jones:2016ldq}.
For exclusive $\Jpsi$ photoproduction, $\mu = \mathcal{O}(m_c)$, where $m_c$ is
the charm quark mass, so such corrections are of $\mathcal{O}(1)$ and are
crucial ingredients in the description of the process. Moreover, the choice $\mu
= m_c$ allows for the resummation of a particular class of large double
logarithms $(\alpha_s \ln(1/\xi) \ln(m_c^2/\mu^2))^n$ that arise in the
high-energy limit of the amplitude~\cite{Jones:2015nna}. Collectively, the implementation of this
low cut-off procedure and program of effective low-$x$ resummation provide for a
more reliable and stable $\overline{\rm MS}$ amplitude to NLO with a reduced
factorization scale dependence, allowing for a sensible comparison to the
experimental data~\cite{Flett:2019pux}.

The central cross section predictions based on input
GPDs constructed from three global PDF analyses via the Shuvaev transform differ
dramatically in the kinematic region accessible to the LHC but are compatible at
HERA energies, see Fig.~\ref{initialstate:fig2} (right). 
This observation, together with the relative size of the input PDF uncertainties
in the low-$x$ and low-$\mu$ domain compared to the experimental uncertainties
and the systematically tamed scale uncertainty discussed above, motivated an
extraction of a low-$x \sim 10^{-3} - 3 \times 10^{-6}$ and low-$\mu^2 \sim 2.4\,{\rm GeV}^2$ gluon PDF over both HERA and LHC kinematic regions via independent
fitting and statistical reweighting approaches using the exclusive $\Jpsi$
photoproduction data gathered from HERA and the LHC~\cite{Flett:2020duk}.
More global impact studies of this data within a larger fitting framework would
be desirable to allow for refined extractions of PDFs and provide novel
constraints at low $x$ and low $\mu$, which would be of practical use to the PDF
fitting groups and to the wider particle physics community.

\subsection{Challenges and opportunities}
Substantial experimental and theoretical progress has been made towards
developing a coherent picture of the gluonic structure of nuclei from forward
particle production and vector meson photoproduction measurements. However, the
proton PDF extractions using HERA and LHCb $\Jpsi$ photoproduction data
discussed above differ dramatically from proton PDF determinations using LHCb
$\Dzero$ meson production data~\cite{Flett:2019pux}. This discrepancy shows that a great deal of work is
needed to reconcile these two classes of observables through better theoretical understanding of the processes e.g.\ in terms of low-$x$ resummations and applied subtraction schemes. This reconciliation is
becoming increasingly urgent as nPDFs achieve new levels of precision yet
continue to produce predictions compatible with CGC calculations, demonstrating
the difficulty of conclusively observing the effects of nonlinear QCD. Vital new
experimental inputs are expected in the near future as the LHC prepares to
produce high-energy OO and pO collisions. Vector meson
photoproduction data from ${\rm O}{\rm O}$ UPCs could be used to further tame
factorization scale uncertainties by constructing ratios with ${\rm PbPb}$
measurements~\cite{Eskola:2022vaf}. Furthermore, forward particle production measurements in pO collisions will probe the $A$-dependence of gluon suppression in nuclei,
potentially revealing the onset of saturation effects at low $x$.

\section{Jet production and properties in pp and in the medium}
\label{sec:jets}
The fragmentation of energetic quarks and gluons in relativistic heavy-ion collisions is modified with respect to its vacuum counterpart due to the presence of a QGP. Hard-propagating partons scatter with the constituents of this color-deconfined medium both elastically and inelastically (see e.g. Ref.~\cite{Apolinario:2022vzg} for a recent review). Inelastic scattering processes trigger additional radiation governed by an emission probability distinct from the standard Altarelli-Parisi (AP) splitting kernels. One of the peculiarities of the medium-induced spectrum is the absence of a collinear singularity. Consequently, medium-induced emissions typically appear at large angles, sometimes even outside the cone of the reconstructed jet, leading to a degradation of the jet energy. This suppression of the jet \pt spectrum in AA collisions with respect to pp collisions has been experimentally confirmed both at RHIC and LHC energies. An in-depth understanding of energy loss mechanisms requires an experimental program that goes significantly beyond inclusive observables. 

During the workshop, we explored the potential of jet substructure measurements to pin down one of the fundamental scales of in-medium jet physics, namely the characteristic angle of medium-induced emissions, $\theta_c$. This angular scale also controls the interference pattern of multiple emitters~\cite{Casalderrey-Solana:2012evi} and, as a consequence, the amount of energy lost. An unambiguous determination of $\theta_c$ would not only inform about these fundamental properties of jet evolution in a dense medium but also about properties of the QGP itself, such as the effective length $L$ or transport properties often characterized by the coefficient $\hat q$, the mean squared \pt transferred per unit length. In this report, we first review the current status of jet substructure measurements that target the determination of $\theta_c$. Next, we discuss the potential of new observables proposed in the talks by the participants and conclude with a brief overview of the follow-up discussions.
\begin{figure*}
  \includegraphics[width=0.32\textwidth]{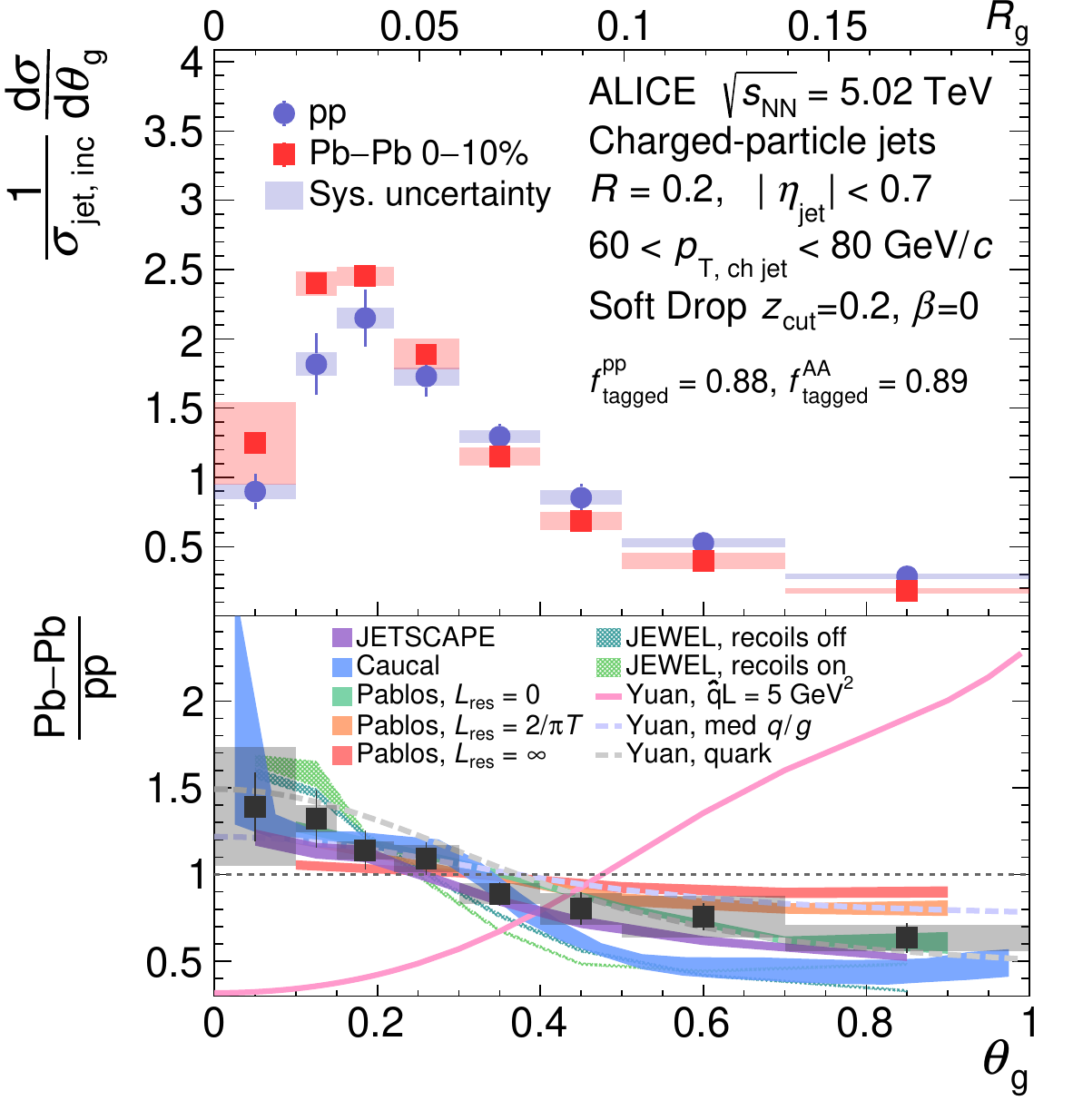}
  \includegraphics[width=0.32\textwidth]{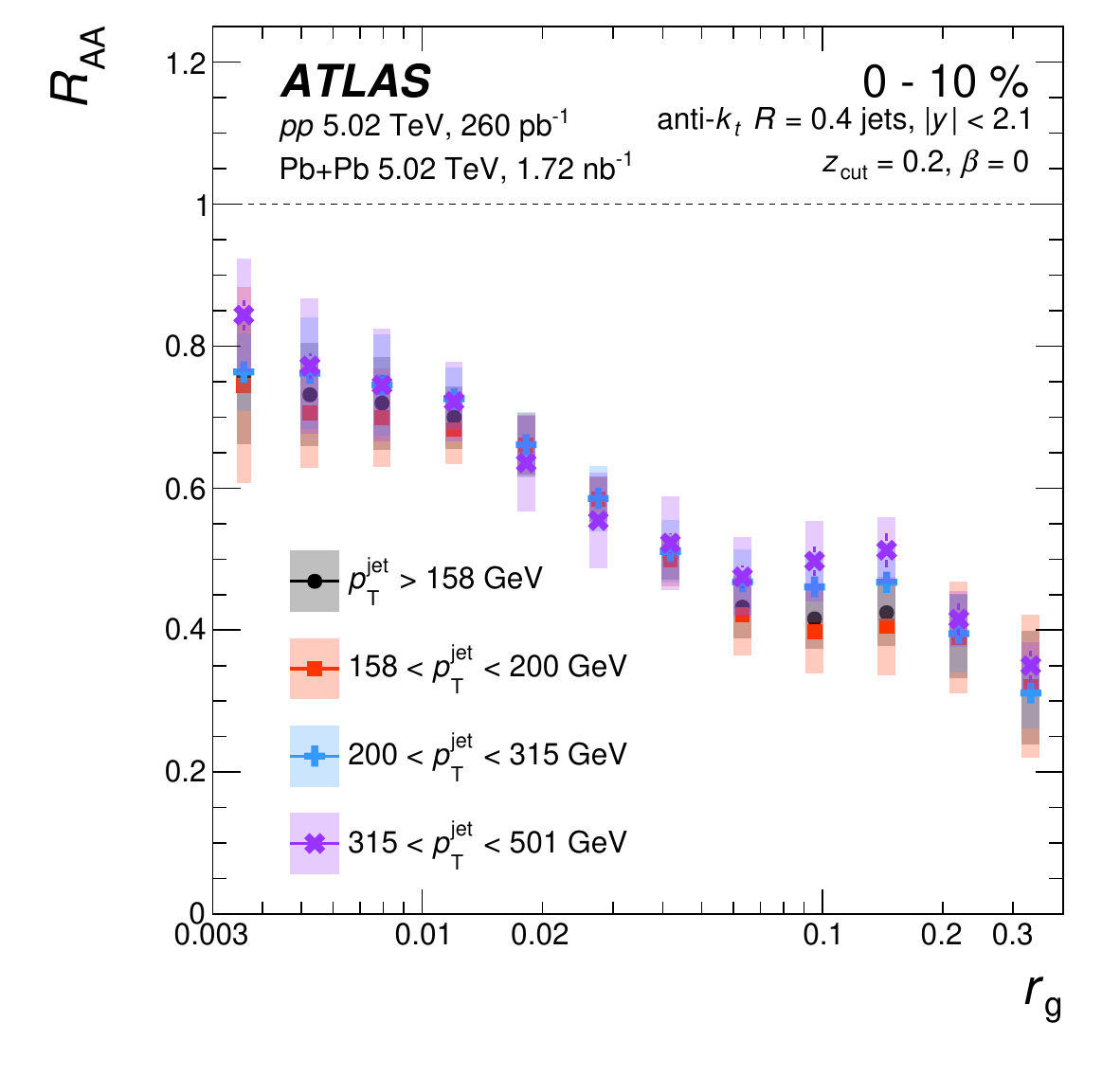}
  \includegraphics[width=0.32\textwidth]{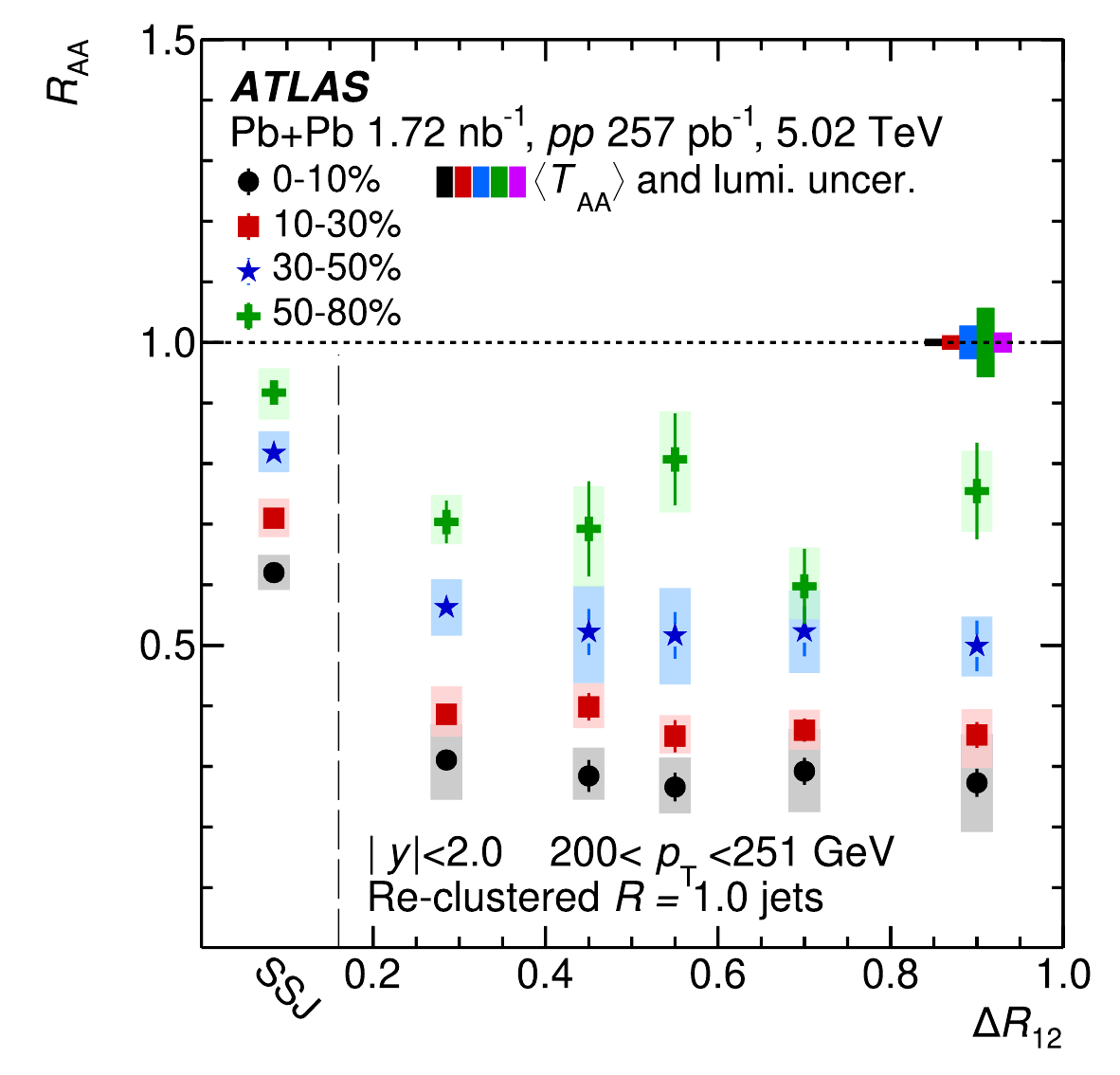}
  \caption{A compilation of recent jet substructure measurements indicating a narrowing of the jet core in heavy-ion collisions. From left-to-right the figures have been extracted from Refs.~\cite{ALargeIonColliderExperiment:2021mqf, ATLAS:2022vii, ATLAS:2023hso}.}
  \label{fig1:jets}
  \end{figure*}
\subsection{Experimental searches of color coherence effects}
The quest for color coherence effects in jet observables has been an active field of research during the last decade. Here, we focus on the most recent developments and refer the reader to Ref.~\cite{Cunqueiro:2021wls} for a complete review. 
One of the most recent measurements aiming at unveiling color coherence dynamics is that of the groomed jet radius, $\theta_g$, by the ALICE experiment~\cite{ALargeIonColliderExperiment:2021mqf} (Fig.~\ref{fig1:jets}, left). The distribution was measured to be narrower in PbPb than in pp, i.e. splittings with $\theta_g<0.2$ are enhanced in the medium. The data points agree with a wide set of theoretical predictions based on quite different ingredients, some of them lacking a color coherence angle implementation.

The ATLAS collaboration extended the measurement of $\theta_g$ by studying the nuclear modification factor $R_{\rm AA}$ as a function of $\theta_g$~\cite{ATLAS:2022vii} (Fig.~\ref{fig1:jets}, middle). For pp collisions, this double differential measurement revealed sizeable discrepancies (up to 50\%) between general purpose Monte Carlo event generators and data, thus highlighting the importance of measuring the pp baseline to guarantee a meaningful interpretation of heavy-ion-to-pp ratios. Regarding the PbPb data, ATLAS observes 
an increasing jet suppression with increasing $\theta_g$, with a possible inflection point at around $\theta_g=0.05$.
This result is in qualitative agreement with a very recent measurement~\cite{ATLAS:2023hso} that does not rely on SoftDrop grooming but rather utilizes a re-clustering procedure in the spirit of trimming~\cite{Krohn:2009th} in which hard subjets are found inside large-$R$ jets. In that measurement (Fig.~\ref{fig1:jets}, right), $R_{\rm AA}$ was evaluated for configurations with different subjet multiplicity, and it was found that large-$R$ jets with single subjets are quenched significantly less than jets with a complex topology (a factor of 2 difference in terms of the jet $R_{\rm AA}$). In addition, $R_{\rm AA}$ remains flat as a function of the opening angle between subjets ($\Delta R_{12}$) for subjets with $\Delta R_{12} > 0.2$ which may be interpreted as a constraint on the maximal value of $\theta_c$. In fact, the value of $\theta_c < 0.2$ is consistent with previous measurements of radius ($r$) dependent fragmentation functions \cite{ATLAS:2019pid} where only minimal changes of the jet fragmentation pattern in PbPb collisions were observed for $r < 0.1$.

The aforementioned measurements point to a narrowing of the angular scale of jets in PbPb relative to pp collision. A possible caveat of such inclusive jet measurements is that the comparison is made at the same reconstructed jet energy, which does not map to the same scattered parton energy in pp and PbPb due to the additional energy loss that is present in the latter case. This can potentially lead to selection biases \cite{Brewer:2021hmh,Du:2020pmp}.

The theoretical interpretation of these results is not yet unambiguous and it is therefore important to discuss novel observables with $\theta_c$-sensitivity that can help disentangle between different effects.

\subsection{Novel observables with $\theta_c$-sensitivity}
The observables discussed during the meeting can be categorized into two classes: clustering-tree based and energy-flow based. In the former category two proposals rely on the dynamical grooming (DyG) method~\cite{Mehtar-Tani:2019rrk}. On the one hand, the dynamically groomed jet radius (akin to $\theta_g$ but probing a wider region of phase-space) was proposed as a potential candidate to detect color coherence effects~\cite{Caucal:2021cfb}. This observable has been already measured in pp collisions~\cite{ALICE:2022hyz} and benchmarked against a theoretical prediction at high-logarithmic accuracy in perturbative QCD~\cite{Caucal:2021bae}. In the medium case, an analytic calculation was also presented together with a dedicated MC study of non-perturbative effects such as hadronisation and medium response. The key difference between the SoftDrop and Dynamical Grooming versions of $\theta_g$ is that while the former leads to a shift in the distribution towards smaller values in PbPb, the latter predicts a genuine change in the shape of the distribution, i.e. the vacuum distribution is almost flat (for some DyG settings) and color coherence dynamics induces a peak around $\theta_c$. The discriminating power of this observable is therefore potentially enhanced. 

Vacuum and medium-modified showers factorize in time to first approximation~\cite{Caucal:2018dla}, with the harder vacuum part happening earlier. A broader early vacuum shower will undergo a more active medium-induced shower due to the larger number of emitters. This introduces a jet-width dependence of energy loss, which is naturally inter wound with flavour dependence.  These effects coexist with color coherence, which further regulates the effective number of emitters in the medium.

The color charge dependence of inclusive jet suppression was explicitly studied in several phenomenological works, see e.g. Refs.~\cite{Spousta:2015fca,Qiu:2019sfj}. An experimental strategy to distinguish between these confounding effects was introduced in Ref.~\cite{Pablos:2022mrx} which proposes to exploit the forward capabilities of LHC detectors.

Finally, a complementary set of observables that do not rely on clustering sequences are the so-called energy-energy correlators~\cite{Basham:1978bw} (EECs) that describe $n$-point correlation functions of energy flux. In recent years, there has been a renewed interest on EECs both theoretically and experimentally. On the formal side, the logarithmic structure of these observables in vacuum entails a beautiful connection to light-ray operators and conformal field theory~\cite{Hofman:2008ar, Kravchuk:2018htv, Dixon:2019uzg}. In a heavy-ion context, first steps towards calculating the two-point energy-energy correlator were presented in Refs.~\cite{Andres:2022ovj, Andres:2023xwr}. A leading-order calculation demonstrated an enhancement of this observable at large angles attributed to color decoherence dynamics. The authors also proposed an experimental measurement of this observable in $\gamma$+jet events to mitigate the selection bias effect. The quantitative impact of energy loss, underlying event and medium response on this observable, although formally power-suppressed by construction of the observable, remains to be studied. It would also be beneficial to perform an in-depth study of the complementarity of clustering-tree based and energy-flow based observables to study jet quenching.

\subsection{Future steps}
The multiple scales involved in the evolution of jets in heavy-ion collisions makes any attempt of isolating $\theta_c$-dynamics an extremely challenging task. We identified a few basic requirements that any observable needs to meet in order to qualify for such a task: (i) calculable vacuum baseline, (ii) resilience to underlying event, and (iii) reduced sensitivity to medium response. We believe that jet substructure observables computed in high-\pt jets could potentially meet these requirements since a sufficiently large separation of scales is guaranteed. 

\section{Event properties and hydro in small and large systems}
\label{sec:hydro}
The development of collective flow behavior \cite{ALICE:2014dwt, ATLAS:2017rtr, CMS:2017kcs, PHENIX:2018lia} and the enhancement of strange hadron production \cite{STAR:2007cqw, ALICE:2016fzo} have been considered as signatures of the presence of the QGP in nuclear collisions.
Such phenomena were considered exclusive to heavy-ion collisions until similar features were also observed in high-multiplicity pp and pA collisions. This raised the question of whether they are due to the formation of a strongly interacting QGP medium or could be explained by other mechanisms, e.g., initial-state momentum correlations. The former scenario is supported by the success of hydrodynamic frameworks in describing the anisotropic flow patterns seen in small collision systems~\cite{Nagle:2018nvi}.
However, collective-like effects have been observed also in ultraperipheral AA collisions \cite{ATLAS:2021jhn} and recent ALICE results suggest that a non zero elliptic flow is measured even in low-multiplicity pp collisions.
Therefore, small systems have grown into prominence in the study of the hot QCD medium and are no longer regarded solely as reference measurements for the studies performed with heavy-ion collisions. This discussion track focused on the frameworks describing event properties and hydro in small and large systems.

\subsection{Global description across systems: core-corona approaches}
\label{sec:corecorona}

The open questions regarding the interpretation of QGP-like effects in small colliding systems require further theoretical development of the description of the medium created in relativistic nuclear collisions. In particular, there is a need for phenomenological approaches 
that take into account non-equilibrium effects that become more and more important going from large to small systems and that also include the description of both a `core' (hydrodynamically evolving QGP) and a `corona' (non-thermalized particles at low densities or high momentum) part of the produced medium~\cite{Werner:2007bf}.

Indeed, the relativistic hydrodynamic approaches describe the matter in local thermal equilibrium (ideal hydro) or with moderate deviations around it (viscous and anisotropic hydro). Still, other parts of the system, such as matter far out-of-equilibrium and propagating jets, are often not included in the hydrodynamic description. This limits the applicability of the hydro framework for small systems and at lower collision energies.
The core-corona approaches try to solve this issue, allowing simulations of large and small systems, as well as a wide range of collision energies: from the highest LHC energies down to the lowest energies of the beam energy scan at RHIC. It is worth mentioning that this capability is also intrinsic in kinetic approaches, in which both the partonic and the hadronic phases are described through the Boltzmann equations (e.g., AMPT~\cite{Lin:2004en}) or generalized transport equations (e.g., PHSD ~\cite{Cassing:2009vt}). 
Here the dynamical separation of a core and corona part is automatically achieved during the space-time evolution of the fireball created in relativistic nuclear collisions. In this sense, such approaches can be considered as belonging to the broader class of core-corona approaches.

Two examples of core-corona models with hydrodynamic formulation for the QGP phase are DCCI2~\cite{Kanakubo:2021qcw, Kanakubo:2022ual}, EPOS2/3 \cite{Werner:2010aa,Werner:2013tya} and EPOS4~\cite{Werner:2023zvo,werner:2023-epos4-micro}. The latter has been recently publicly released~\cite{EPOS4}. The experimental data on charged-particle multiplicity and $\Omega/\pi$ ratio from pp to AA collisions are used both in the DCCI2 model and EPOS to fix the model parameters.



\begin{figure*}
\centering
\includegraphics[width=0.33\textwidth]{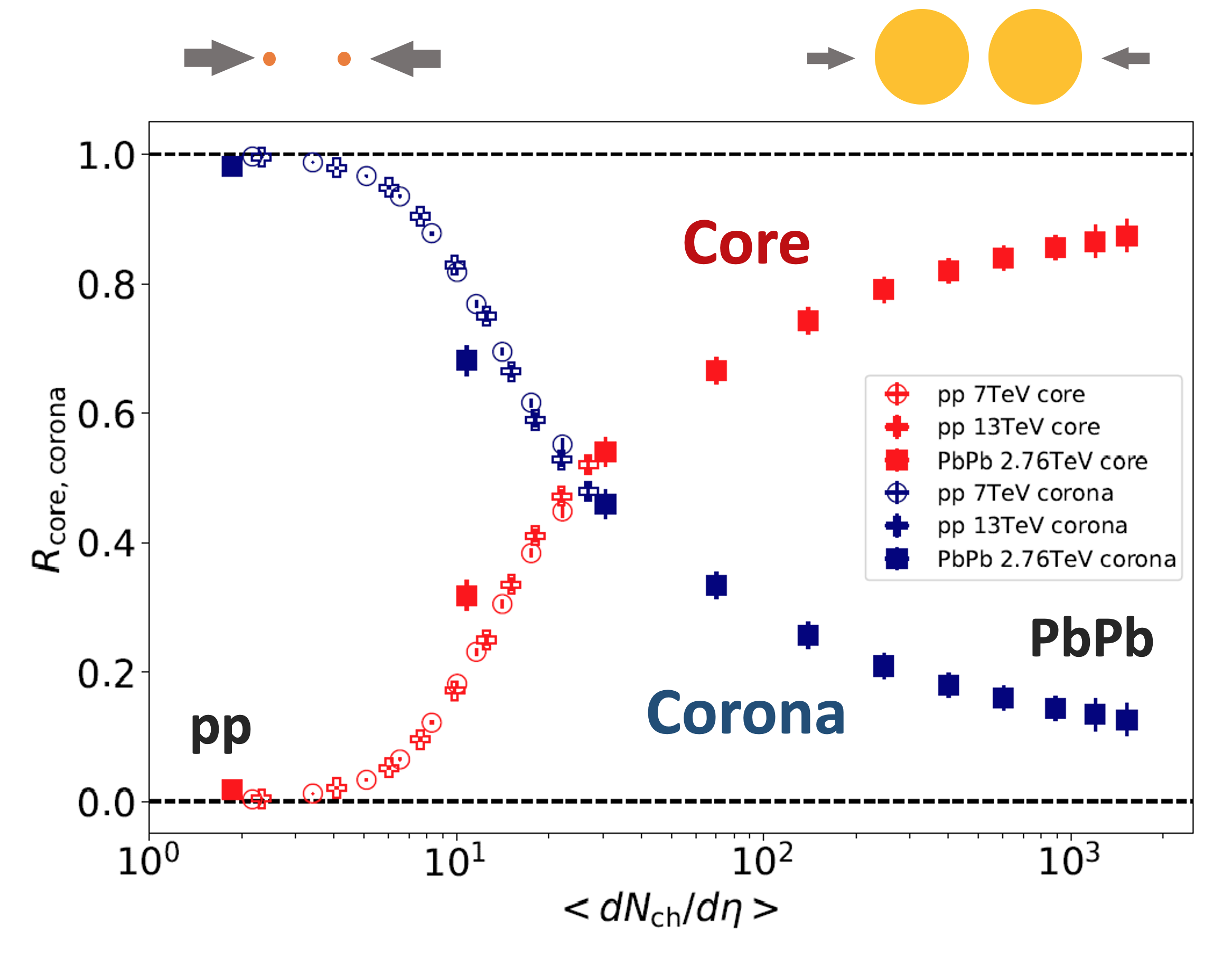}
\includegraphics[width=0.64\textwidth]{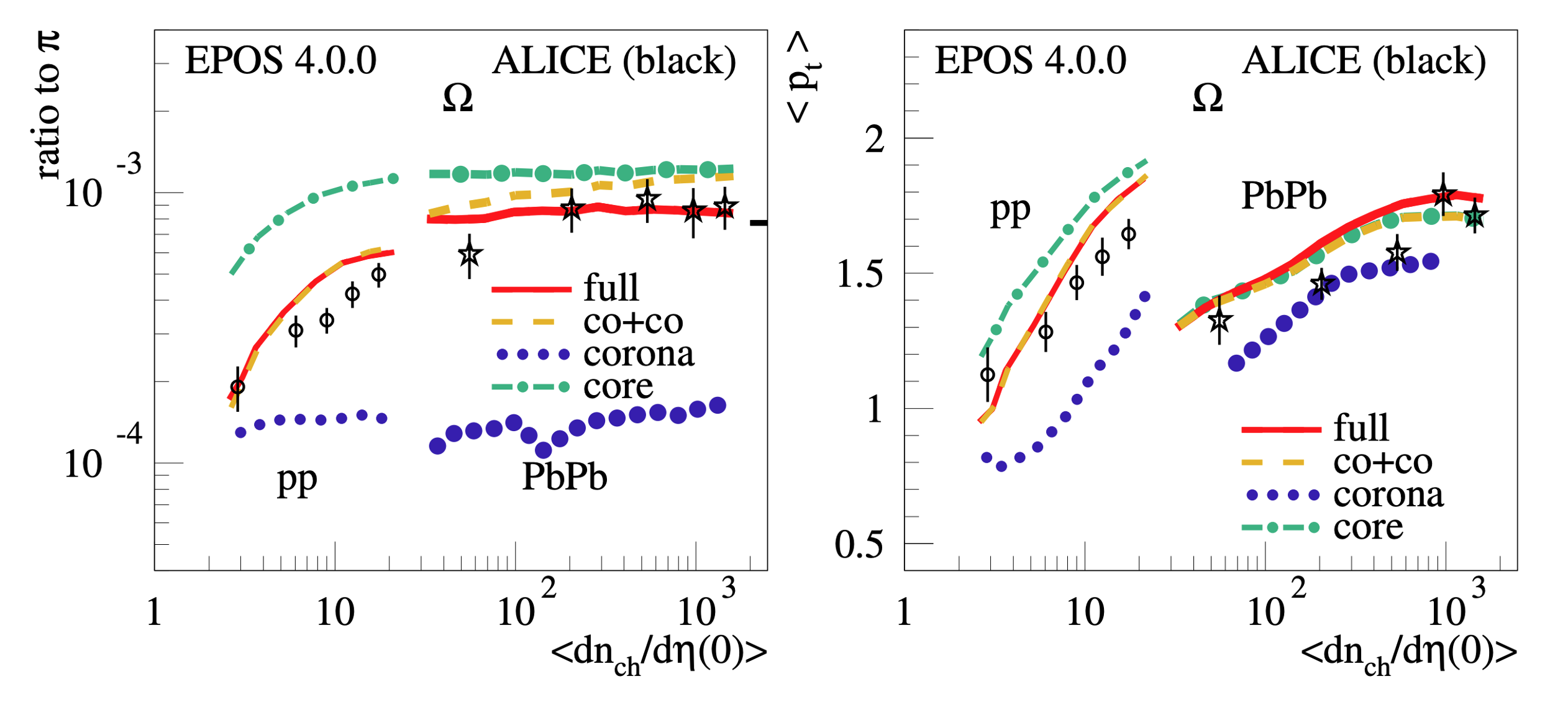}
\caption{Left: fraction of core and corona as a function of multiplicity from pp to PbPb. Middle and right: integrated $\Omega/\pi$ yield ratio (middle) and average transverse momentum (right) as a function of charged-particle multiplicity density at midrapidity from pp to PbPb compared to EPOS 4.0.0 predictions from various parts of the collision system~\cite{Werner:2023jps}. The \enquote{co+co} (core+corona) curve is an interpolation between the core and corona cases, with the core weight increasing continuously with multiplicity. The \enquote{full} case is equal to the co+co case with in addition the inclusion of hadron rescattering.}
\label{fig:hydro-fig1}
\end{figure*}

As seen from Fig.~\ref{fig:hydro-fig1} (left) the fractions of the core and corona parts exhibit a clear scaling with multiplicity.
There is a smooth increase of the core part and a decrease of the corona part going from low to high multiplicity.
The onset of core dominance within the DCCI2 approach happens at $\langle dN_{ch}/d\eta\rangle\sim 20$.
The same scaling is present in observables sensitive to particle multiplicity, such as the $\Omega/\pi$ ratio, see Fig.~\ref{fig:hydro-fig1}(middle).  EPOS can also describe the trend of the experimental measurements for the mean transverse momentum $\langle \pt \rangle$ as a function of multiplicity shown in Fig.~\ref{fig:hydro-fig1} (right), including the jump seen when passing from pp to PbPb collisions.
Therefore, core-corona approaches are able to describe the hadrochemistry and  kinematics across systems. It is interesting to note that the core part has an important role even in pp collisions. 

Furthermore, an important feature of core-corona models is the ability to describe multiple observables across many collision systems simultaneously. This feature is crucial for an over-arching understanding of how similar phenomena can appear from small to large collision systems. 

\subsection{Deeper insights into nuclear collisions: realistic 3D view}
While the majority of studies are related to observables at midrapidity, in the last decade, it has become more and more important to gain a realistic three-dimensional view of the entire interaction area of the collision. For example, describing the tilted fireball produced in non-central heavy-ion reactions \cite{Bozek:2010bi} as well as the asymmetric particle rapidity distributions generated in pA and other asymmetric collisions \cite{Zhao:2022ugy, Oliva:2019kin}.
In relativistic AA collisions, the tilt of the source is converted by the collective expansion of the produced hot matter into the directed flow $v_1$, which is the first harmonic of the particle azimuthal distribution in momentum space. The directed flow of light hadrons constitutes a probe of the asymmetric geometry of the matter distribution in the longitudinal direction in pA collisions~\cite{Oliva:2019kin}. 
Full 3D simulations are also crucial for the comparison to the elliptic and triangular flow data in small systems. In particular, in~\cite{Zhao:2022ugy}  it is estimated that approximately 50\% of the difference between triangular-flow values measured by PHENIX and STAR originates from the the different flow correlations between different rapidity regions. This highlights the importance of the correct 3D view of the fireball and requires theoretical approaches to have the initial condition modeled in ‘realistic’ ways and support further investigation of collectivity and correlations at large rapidity.

Surprisingly, it turned out recently that the $v_1$ of heavy-flavor particles is much larger than that of light charged hadrons in both large and small systems \cite{Chatterjee:2017ahy, Haque:2021qka}; the origin of the $v_1$ of D mesons is different from the one of the bulk medium and its large value manifest only due to the non-perturbative interaction of heavy quarks with the QGP medium \cite{Oliva:2020doe}. Therefore, the directed flow allows 3D access to the production of both soft and hard probes and a deeper look into their mutual interaction.

LHCb collaboration is a relative newcomer to collective flow measurements in heavy-ion collisions. However, thanks to upgrades in Run 3 LHCb will be able to study PbPb collisions up to 30\% centrality. The large forward rapidity reach of LHCb puts it in a unique position to contribute to the study of the full 3D picture of nuclear collisions. In parallel, LHCb will take data in a fixed target mode with SMOG2 for a wide range of nuclear targets at a smaller center of mass energy but with high luminosity, allowing to probe a variety of initial-state geometries. 

\subsection{The evergreen tale of collectivity}

Multi-particle correlation measurements are arguably the most versatile and precise tool in studying collective phenomena in nuclear collisions. The centrality and momentum dependence of elliptic and triangular flow coefficients are the key inputs of the state-of-the-art Bayesian inference analyses of QGP properties like specific shear and bulk viscosities~\cite{Bernhard:2019bmu,JETSCAPE:2020mzn,Nijs:2022rme}. The momentum dependence of baryon and meson elliptic flow is used to benchmark different hadronization mechanisms (e.g. fragmentation or coalescence) at different momentum scales. More generally the observed similar mass ordering of collective flow in PbPb and pPb collisions points to the common origin of this effect. 

\begin{figure}[hb!]
\centering
\includegraphics[width=0.45\textwidth]{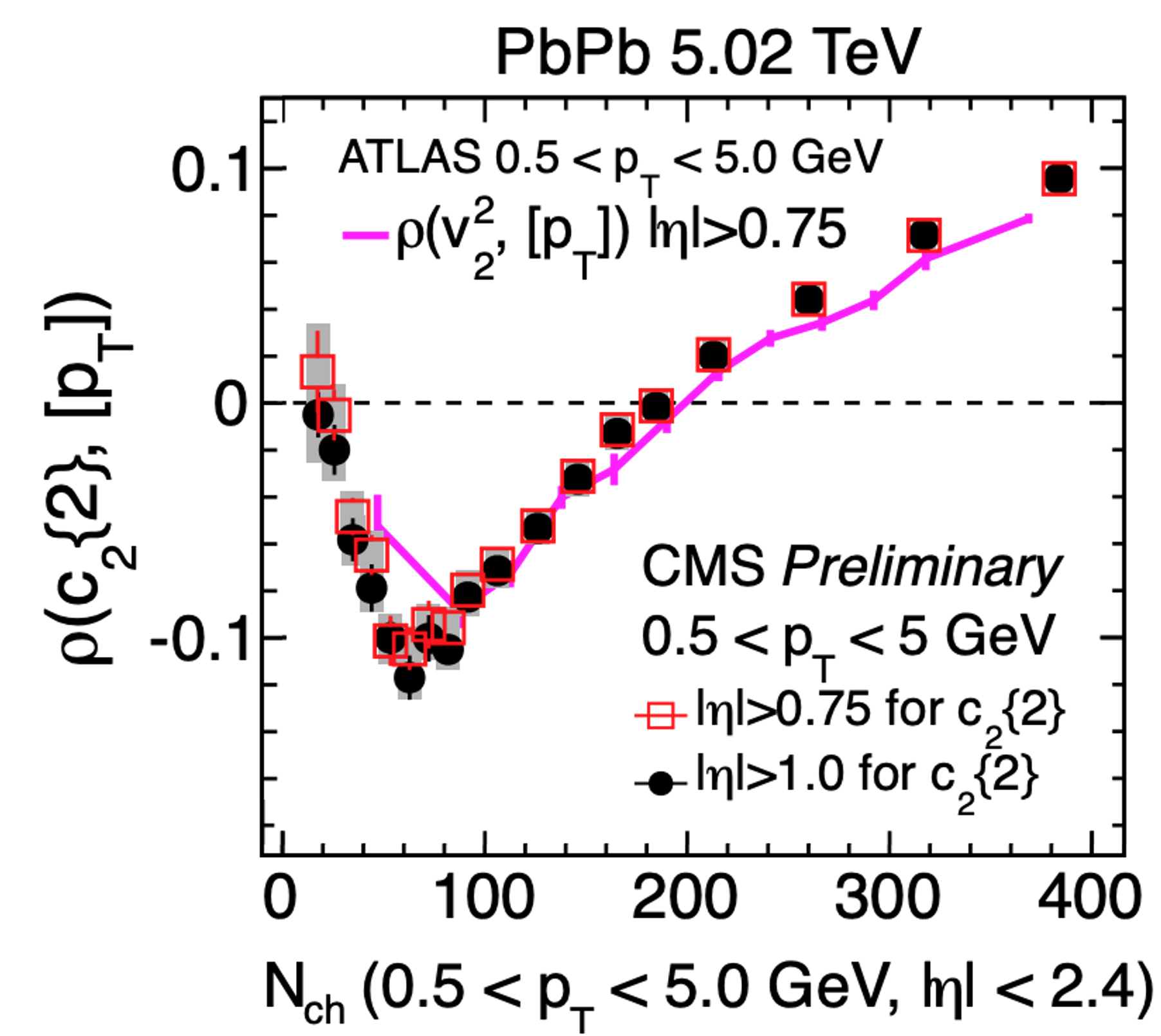}
\caption{Pearson correlation coefficient $\rho(c_{2}\{2\}, [p_{T}])$ as a function of multiplicity as measured by ATLAS~\cite{Aad_2019} and CMS~\cite{CMS-PAS-HIN-21-012}.}
\label{fig:hydro-fig2}
\end{figure}

The crucial outstanding question is whether the initial state effects, e.g. initial momentum correlations, could be resolved in flow measurements in small systems. One very promising observable is the correlation between fluctuations in mean \pt and elliptic flow magnitude~\cite{Mazeliauskas:2015efa, Bozek:2016yoj}. In central PbPb collisions, the elliptic and radial flows have a positive correlation. When inspecting this correlation as a function of multiplicity, it is important to consider that the correlation between particle multiplicity and impact parameter becomes weaker going towards peripheral collisions, and it is even more diluted in small collision systems (see e.g.\cite{Oliva:2019kin}).
For a given multiplicity interval, at a larger impact parameter (smaller overlap area) the initial energy density must increase to maintain the same multiplicity. In very peripheral collisions, the collision geometry is no longer controlled by the impact parameter, but rather by the large fluctuations of the number of participating nucleons relative to the average. With the net entropy results, more deformed shapes have a lower density than more compact configurations leading to the anti-correlation of elliptic flow and mean \pt fluctuations. In contrast, the initial state models predict positive correlations. Indeed the Pearson correlation coefficient measured by ATLAS and CMS collaborations shows the characteristic change in behavior at low and high multiplicities, as shown in Fig.~\ref{fig:hydro-fig2}. However, a firm conclusion about measuring the initial state effects is complicated by the measurement's sensitivity to experimental cuts and similar behavior seen in models without initial state correlations, e.g. AMPT~\cite{Lim:2021auv}. Further studies of this fascinating behavior will be pursued.

The difficulty to link the collective behavior seen in small systems to initial-state effects or to identify it as a QGP signature has stimulated first attempts to study observables related to particle production and collectivity by means of multidifferential analysis, in which centrality selection is flanked by event-shape categorization of events.
An example of such event-shape selection methods employed from small to large systems is the transverse spherocity analysis~\cite{Banfi:2010xy, Ortiz:2015ttf, ALICE:2019dfi, Mallick:2020ium, Oliva:2022rsv}, which has been used to classify events based on different degree of the collective effects in heavy-ion collisions \cite{Mallick:2020ium}. Other event-shape selection techniques are those based on the relative transverse activity classifier \cite{Martin:2016igp, Bencedi:2021tst} and on flattenicity \cite{Ortiz:2022zqr, Ortiz:2022mfv}, mainly employed in pp collisions, as well as the event-shape engineering based on flow vector~\cite{Poskanzer:1998yz}, until now exploited especially in heavy-ion experiments at LHC for studying correlations between flow harmonics of different order~\cite{ATLAS:2015qwl}.
Various studies are exploring the potentiality of these event-shape classifiers. For instance, the capability of flattenicity to identify collisions with multiple lower transverse-momentum parton-parton scatterings and softer \pt spectrum compared to the multiplicity estimator may be exploited during LHC Run 3 and 4 to isolate high-multiplicity pp collisions originated by soft partonic processes. Moreover, flattenicity is a promising variable to overcome the drawback of multiplicity-based event classifiers that, suffering from biases towards multijet final states, make challenging the quest for jet-quenching in high-multiplicity pp collisions~\cite{Ortiz:2022mfv}.
Pursuing further studies based on multidifferential measurements in pp and in pA collisions may help to gain deeper insights into the origin of collectivity in small colliding systems.

The collective effects of QGP can manifest themselves not only in the kinematic distribution of final state particles but also leave imprints in their polarization. The short, but rapid rotation of the medium achieves very large values of vorticity. In a thermal system, this leads to spin alignment of quarks and consequently the spin of produced hadrons.  By measuring the spin of produced hadrons with respect to the collision axis, the amount and direction of hadron polarization can be estimated. Polarization physics has attracted a lot of attention from both experimental and theoretical communities~\cite{Becattini:2022zvf}. The $\Lambda$ polarization puzzle and recently measured vector meson polarization have led to new theoretical developments. 
Various phenomena can be studied with polarization, ranging from global features due to the collectively expanding medium to more localized 
vorticity that appears due to high-energy particles propagating through the medium \cite{Serenone_2021, ribeiro2023lambda}. Notably, vorticity - and subsequently also polarization - has been shown to encode information about medium properties like viscosity, potentially rendering polarization measurements a fundamental tool in characterizing the QGP. 

\section{Hadronization of light and heavy flavour across collision systems}
\label{sec:hadronization}

By now, there is abundant evidence that many non-perturbative ``constants'' such as baryon-to-meson ratios, strangeness fractions, and (moments of) hadron kinematic distributions, can depend sensitively on the type of colliding beam particles (the collision system) and on event activity (the production environment). 

Using rapidity density of charged particles as a proxy for event activity, many (though not all) salient observables exhibit continuous increases with progressively higher average multiplicity densities in $\pp$, $\pA$, and $\AA$ collisions, typically reaching approximately constant asymptotic values for (central) $\AA$ collisions. 

Several physical models have been proposed that are capable of modelling these trends qualitatively, and increasingly even quantitatively. During the workshop the following types of models were discussed: 
\begin{itemize}
    \item General-purpose Monte-Carlo event generators (such as PYTHIA~\cite{Bierlich:2022pfr} and HERWIG~\cite{Bellm:2019zci}) mainly focus on \ee, \ep, and \pp collision systems. Colour reconnections (CR) can increase baryon-to-meson ratios~\cite{Christiansen:2015yqa,Gieseke:2017clv} and non-perturbative colour interactions (CI) between strings or clusters can increase both baryon and strangeness fractions~\cite{Bierlich:2014xba,Fischer:2016zzs,Duncan:2018gfk,Bierlich:2022ned}. Both types of effects (CR and CI) depend (implicitly or explicitly) on the effective local density of coloured partons. Since there is no ``medium'' in these models, the main physical driver for the local density of coloured partons in hadron-hadron collisions is multi-parton interactions (MPI). These are absent in \ee and \ep collisions. Note however that $\ee \to \mathrm{WW} \to \mathrm{hadrons}$ furnishes a case similar to that of double-parton interactions, and was instrumental to demonstrating the existence of CR at LEP~2~\cite{ALEPH:2013dgf}, albeit with relatively low statistics.
    Both mechanisms (CR and CI) can also produce flow-type effects~\cite{OrtizVelasquez:2013ofg,Bierlich:2016vgw,Duncan:2019poz,Bierlich:2020naj}. 
    
    \item Models of quark coalescence are based on the idea that in the dense medium of quarks and gluons that is formed in high-energy AA collisions, but even in high multiplicity pA and pp ones, hadrons come from the coalescence of quarks with a similar velocity. Nearly twenty years ago it was realized that such an hadronization mechanism would be quite efficient in creating baryons, resulting in larger baryon yields w.r.t. those expected from standard fragmentation functions. Most of the coalescence approaches evaluate the hadrons yields as a convolution of the quark thermal (at least at low \pt) distribution functions at a temperature $T_H \simeq 155\, \rm MeV$ and the hadron wave function \cite{Greco:2003mm,Fries:2008hs}. A statistical homogeneous color distribution is assumed in evaluating the probability of quark coalescence. 
    An alternative formulation of the idea of recombination instead evaluates the hadron formation as a convolution between the quark distribution function with a Breit-Wigner resonant scattering cross section instead of the hadron Wigner wave function \cite{Ravagli:2007xx}. Despite differences in the details, all these approaches show the general feature of an enhancement of baryon to meson ratio at intermediate \pt with $\mathrm{p}/\pi$, $\Lambda/\kzero \simeq 1$ \cite{Fries:2008hs}; furthermore they all agree that at $\pt > 6-8 \,\rm GeV$ the dominant process will be independent string fragmentation. Coalescence models share also the feature of predicting a quark-number scaling of the hadron elliptic flow, which has been also observed experimentally \cite{Greco:2003mm,Fries:2008hs,Minissale:2015zwa}.
    In the last decade these approaches have been extended to the charm sector and have successfully predicted large values of the $\Lc/\Dzero$ ratio~\cite{Oh:2009zj,Plumari:2017ntm} with peak reaching about unity in AA collisions and about 0.5 in pp and pA collisions~\cite{Minissale:2020bif}.
    
    \item Statistical hadronization models  (SHM) are successful in describing hadron yields in (central) heavy-ion collisions over a broad range of collision energies~\cite{Andronic:2017pug}. The evolution of hadron ratios with multiplicity in pp, pPb and  collisions is also well described~\cite{Cleymans:2020fsc}, here with some assumptions on the canonical conservation volume for strangeness and baryon numbers~\cite{Vovchenko:2019kes}.
    \item Core-corona models are designed to simulate relativistic pp, pA, and AA collisions across a wide range of energies. These models aims to provide a unified approach by considering collective effects in all systems and assuming the creation of a medium. In the core-corona procedure, high-multiplicity events are made by a dense core (bulk-matter) that thermalizes and expands collectively, and a corona near the surface. The separation between core and corona is considered a dynamical process based on density. The two main models considered are EPOS and DCCI and are explained in detail in Sec. \ref{sec:corecorona}.
    
    \item  Another approach to in-medium hadronization, also based on the idea of a quark recombination, is the one developed on the background of a POWHEG+PYTHIA Monte Carlo event generators~\cite{Beraudo:2023nlq}. On a hadronization surface identified by the local temperature $T_H=155\, \rm MeV$, a charm quark is recombined  with a thermal quark or di-quark at finite mass values typical of those of the constituent quark picture. The recombination occurs if the invariant mass of the pair, $M_{\cal C}$, is larger than the mass of the lightest charmed hadron. Then the cluster decays into a charmed hadron that has the same baryon number and strangeness of the cluster or, for large invariant masses, $M_{\cal C}> M_{max}\simeq 4 \,\rm GeV$, it hadronizes via string-fragmentation according to PYTHIA 6.4. 
    A specific trait of this approach is to assume a thermal population of di-quarks that acquire also a hydrodynamical radial and elliptic flows. In Sect.\ref{sec:heavyflav}, this feature will be discussed in relation to the elliptic flow of $\Lc$.
    
\end{itemize}
In Sect.~\ref{sec:lightflav} and~\ref{sec:heavyflav} respectively, the current status in terms of available experimental measurements and their description by theoretical models is presented, focusing on light and heavy flavours, respectively.
In Sect.~\ref{sec:exoticHad}, the current status of observations and physical models of exotic hadrons (carrying charm quarks) is discussed. The structure of these hadrons, whether tetraquarks or pentaquarks, is interesting and currently theoretically debated. These hadrons can also shed light on hadronization mechanisms, as suggested by the recent observation by LHCb \cite{LHCb:2020sey} of an event-multiplicity dependence of the $\chi_{c1}(3872)$ hadron production in pp collisions.  
Finally, in Sect.~\ref{sec:hadNext}, a summary is followed by a menu for next steps. 

\subsection{Light Flavours \label{sec:lightflav}}
One of the most intensely studied observable is the multiplicity dependence of the yield ratio to pions for various strange particle species. Measured by ALICE~\cite{ALICE:2016fzo} across $\mathrm{pp}$, $\mathrm{pA}$, and $\mathrm{AA}$ collisions, the ratio is observed to exhibit a strong multiplicity dependence for $\mathrm{pp}$ and $\mathrm{pA}$ collisions, progressively stronger for multi-strange hyperons.

It was discussed during the workshop that it would be advisable to indicate also the values measured in $\mathrm{e^+e^-}$ collisions when showing these plots, for a clearer ``vacuum'' reference value. Interesting attempts were also made to separate this into quark and gluon jets~\cite{OPAL:1998izp} in $\mathrm{e^+e^-}$ collisions: the interpretation of the data is however complex also because of the dependence of the results on the details of the method used to separate the classes. Non-trivial multiplicity dependence of these quantities in $\mathrm{e^+e^-}$ collisions is also possible~\cite{Hunt-Smith:2020lul}, though so far little explored.

\begin{figure*}[htb]
\begin{tabular}{ll} \begin{minipage}{.58\textwidth}
\hspace{-.7cm}\includegraphics[width=1.\textwidth]{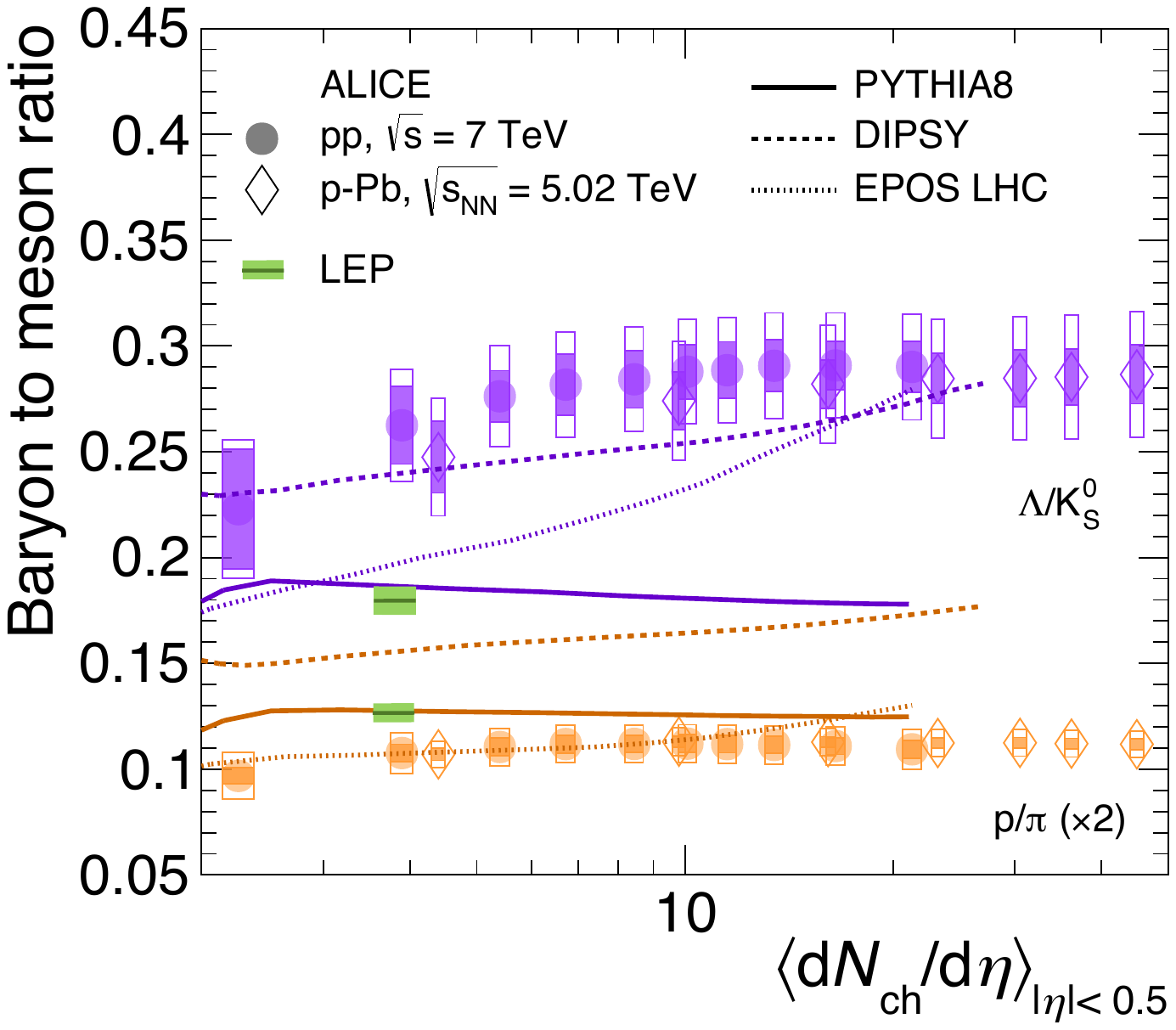}
  \end{minipage} & \begin{minipage}{.42\textwidth}
\vspace{-.6cm}
\hspace{-.9cm}\includegraphics[width=1.02\textwidth]{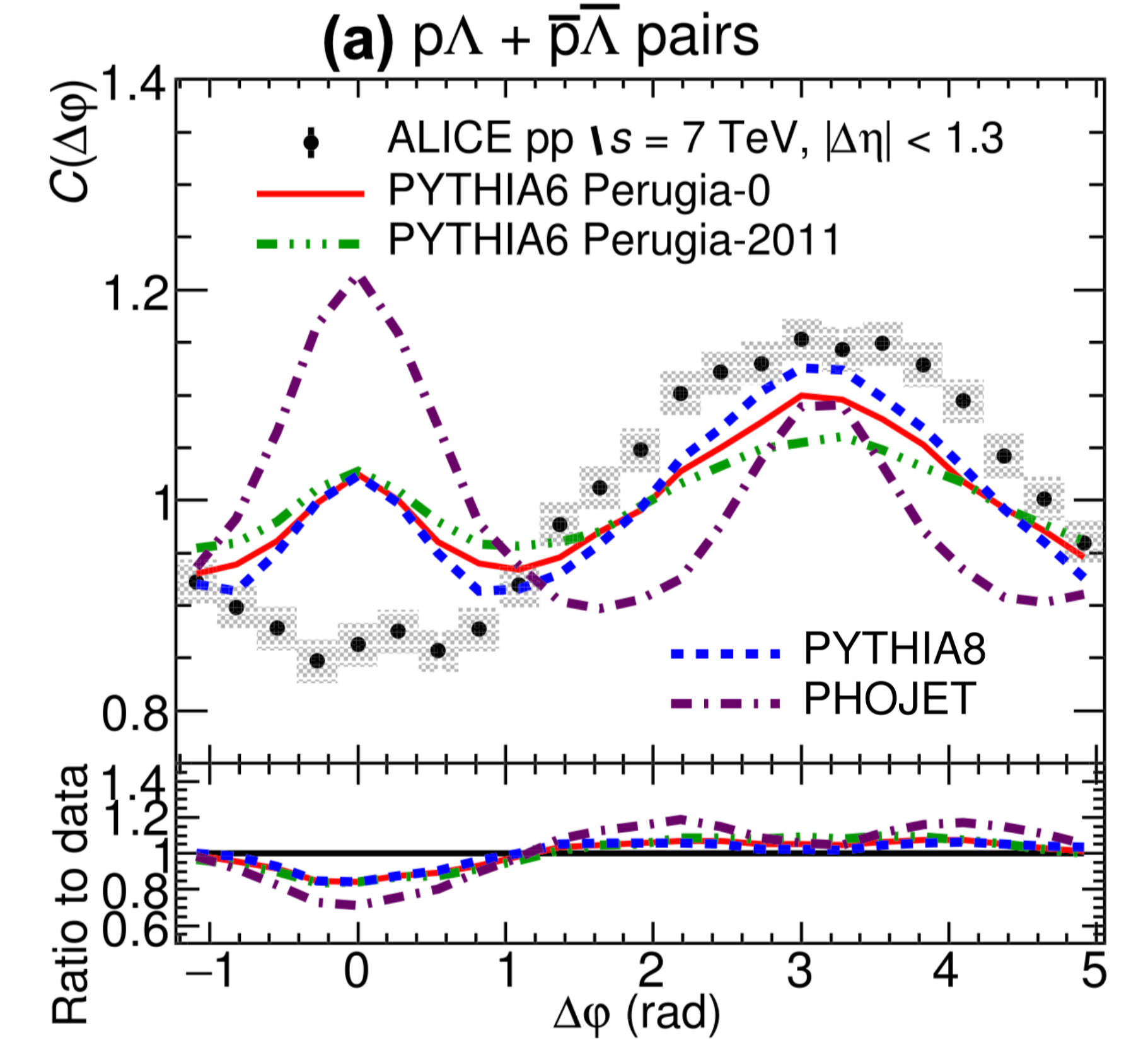}
\end{minipage} \end{tabular}
\caption{Left: Baryon-to-meson ratios measured by ALICE~\cite{ALICE:2016fzo}, with approximate indications of the values measured at LEP (DELPHI~\cite{DELPHI:1996sen}) superimposed. Right: proton-$\Lambda$ correlations in azimuth \cite{ALICE:2016jjg}} \label{fig:had1}
\end{figure*}

An important conundrum that any physical model of these effects will need to address is the fact that the proton-to-pion ratio has been observed to exhibit very little (if any) dependence on multiplicity, combined with the intriguing fact that the values of this ratio observed in LHC collisions appears to be a bit \emph{below} the value observed at LEP, see Fig.~\ref{fig:had1} (left).
According to the PDG table of average identified-particle multiplicities in $\mathrm{e^+e^-}$ collisions~\cite{ParticleDataGroup:2020ssz}, $\left<N_p\right>/\left<N_\pi\right>$ at LEP is about $0.062 \pm 0.002$, whereas the values measured by ALICE in $\mathrm{pp}$ collisions~\cite{ALICE:2016fzo} range from about $0.048 \pm 0.006$ for the lowest charged-particle densities to about $0.055 \pm 0.005$ for the highest ones. 
This presents a challenge for many of the current dynamical models on the market, for which the LEP value typically acts as an effective lower bound. Even in models (or tunes) that do not use the LEP value as a direct constraint, one would still have to address at least in principle what physical mechanism accounts for the universality breakdown between  $\mathrm{e^+e^-}$ and $\mathrm{pp}$ collisions. One potential mechanism that has been hypothesised to possibly reduce the number of proton--antiproton pairs is hadronic re-annihilation~\cite{Sjostrand:2020gyg} though it may be doubtful if that could be sufficiently active even at low multiplicities, as required by data. Another point that was made during the workshop is that it may be useful as a cross check to study how the numerator and denominator, i.e. the proton and pion yields, evolve separately, although the denominator is of course largely degenerate with the charged multiplicity itself. Other potential cross checks could include measurements of this (and other) ratios in high-$p_\perp$ (quark and gluon) jets. Hadrochemistry in jets is since recently being investigating with data already existing on hyperons~\cite{ALICE:2022ecr} and on deuterons \cite{ALICE:2022ugx}. The former were important to demonstrate and quantify the different production rates of strange baryons relative to $\kzero$ mesons in and out-of jets. The measured ratios also set constraints to di-quark formation in jets to models like PYTHIA8, which cannot reproduce the data. The deuteron measurement instead supports the formation of deuterons in-jets via hadronic interactions occurring between hadrons close in phase-space data. Within uncertainties, the data can be reproduced both by modelling these interactions as a simplistic coalescence process and by a more realistic, four-momentum conserving, hadron rescattering model introduced in PYTHIA8.3.

A further interesting observation in baryon production in $\mathrm{pp}$ collisions is that there is a dip in both baryon-baryon and baryon-antibaryon correlations near $\Delta \phi \sim 0$ \cite{ALICE:2016jjg}, which is not described by MC models, Fig.~\ref{fig:had1} (right), see also ref. \cite{ALICE:2023asw}. The suggestion was made that this might be interesting to follow up with similar measurements involving heavy-flavour baryons, though admittedly experimentally very challenging.

\subsection{Heavy Flavours\label{sec:heavyflav}}

A rather surprising observation at the LHC is the large fragmentation fraction for charmed baryons in pp collisions~\cite{ALICE:2021dhb}, which is significantly larger than that measured at LEP. PYTHIA (CR Mode 2)~\cite{Christiansen:2015yqa} describes $\Lc$ data but underpredicts $\XicPlusZero$ data.
In a statistical hadronization approach, He and Rapp~\cite{He:2019tik} assumed, under the guidance of the Relativistic Quark Model~\cite{Ebert:2011kk}, the existence of a vast set of excited charm-baryon states, most of which yet unobserved~\cite{ParticleDataGroup:2020ssz}, in order to describe the measured $\Lc/\Dzero$ ratio. This approach underpredicts $\XicPlusZero$. SHM was recently applied to bottom hadronization in pp collisions~\cite{He:2022tod}.

\begin{figure*}[hbt]
\centering
\includegraphics[width=0.9\textwidth]{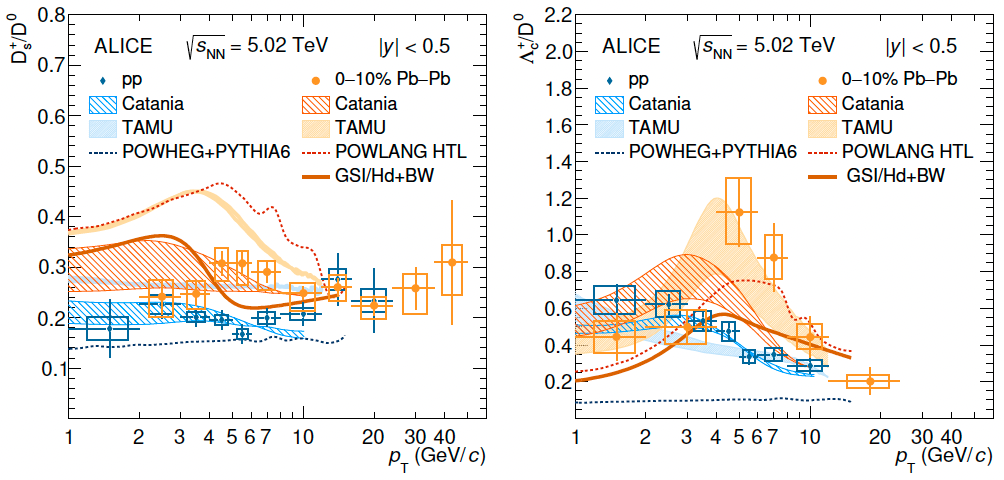}
\caption{Charm hadron ratios vs. \pt, data in comparison to models \cite{ALICE:2022wpn} (Catania: coalescence; TAMU, GSI-Hd: SHM, TAMU with enhanced charm-baryon spectrum, GSI-Hd with the PDG one).}
\label{fig:had2}
\end{figure*}

The current status of the comparison of data and models is shown in Fig.~\ref{fig:had2}, where the ratios of $\Ds$ mesons and $\Lc$ baryons to $\Dzero$ mesons are shown as a function of \pt for pp and central \PbPb collisions. The data exhibit a clear difference between pp and \PbPb, which the models capture well, though overall the model description of the data needs further improvements.

\begin{figure*}[hbt]
\begin{tabular}{ll} \begin{minipage}{.55\textwidth}
\hspace{-.7cm}
\includegraphics[width=1.\textwidth]{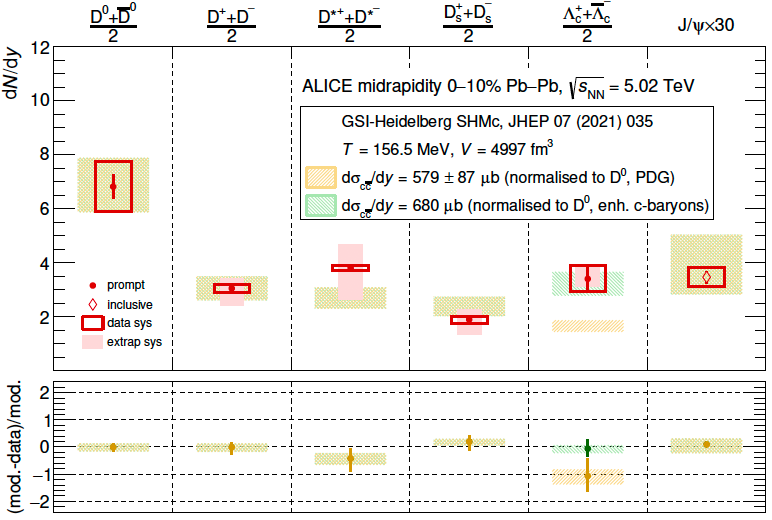}
  \end{minipage} & \begin{minipage}{.45\textwidth}
\vspace{.2cm}
\hspace{-.9cm}\includegraphics[width=1.\textwidth]{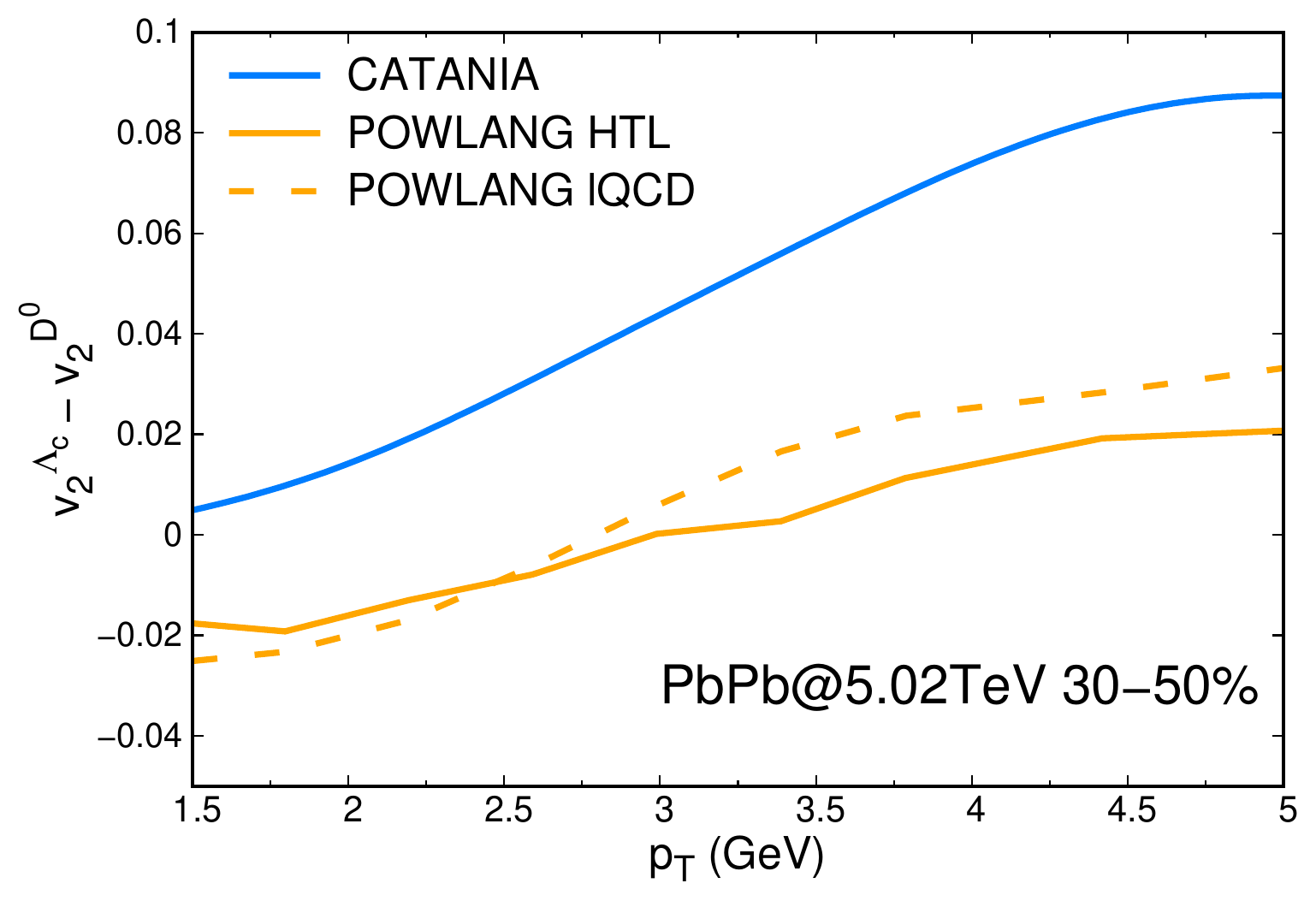}
\end{minipage}
\end{tabular}
\caption{Left: charm hadron yields in central \PbPb collisions, data \cite{ALICE:2022wpn} and SHM predictions. Note the scale factor of 30 for the $\Jpsi$ meson. Right: The difference between the $v_2$ values of $\Lc$ and $\Dzero$ as a function of \pt as predicted by the Catania coalescence model and POWLANG.}
\label{fig:had3}
\end{figure*}

The rather complete charm chemistry was recently measured by ALICE in central \PbPb collisions \cite{ALICE:2022wpn}, see Fig.~\ref{fig:had3} (left). The SHM describes (predicted~\cite{Andronic:2021erx}) the data very well, except for the $\Lambda_c$ baryon, where a good description is achieved only by the inclusion of many (tripled in number) excited charm-baryon states compared to the default (PDG) hadron spectrum. It is important to recall that SHM assumes a concurrent hadronization of all flavors in QGP at the crossover temperature of 156.5 MeV. A similarly quantitative description of charmed hadron yields in dynamical models remains challenging, but developments are under way. Measurements of flow for different charm-hadron species can provide an important test for the modelling of hadronisation. Figure~\ref{fig:had3} (right) shows the \pt dependence of the difference between $v_2(\Lambda_c)$ and $v_2(D^0)$ predicted in a coalescence approach and in the POWLANG mechanism. In the coalescence model the $v_2$ of $\Lambda_c$ receives a contribution from two light quarks while the D meson only from one. In the POWLANG approach instead $\Lc$ combines to a diquark evolving hydrodynamically with the bulk medium and has a milder difference w.r.t. D in particular at intermediate \pt.

In closing this section it is worth mentioning the possible presence of a persisting problem in the data on $\Lc$ baryon, where the rapidity distribution with LHCb (forward rapidity) and ALICE (midrapidity) data appears too peaked at midrapidity.
It is also worth recalling that hadronization in the charm sector is not only interesting per se, but crucial for the attempts to extract the heavy quark transport coefficients via (transport) model description of the data. Currently, the spread in the model predictions due to hadronization is large \cite{Rapp:2018qla}.

\subsection{Exotic Hadrons \label{sec:exoticHad}}

Exotic hadrons are those hadrons which do no fit in the standard meson and baryon classification~\cite{Brambilla:2019esw}. We discuss briefly the charm-carrying tetraquark candidates 
$\chi_{c1}(3872)$, which is charmonium-like (hidden-charm) and the recently-discovered $T_{cc}^+$, a doubly-charmed tetraquark candidate with a mass of about 3875 MeV/$c^2$ and a $\mathrm{cc\bar{u}\bar{d}}$ quark content \cite{LHCb:2021vvq}.
Both hadrons can be interpreted as (loosely-bound) mesonic molecular states, namely $\mathrm{D}^{*0}\bar{\mathrm{D}}^0$ and $\mathrm{D}^{*+}\mathrm{D}^0$, respectively, as their masses lie narrowly below the thresholds for the respective channels. In such an interpretation these hadrons are consequently extended objects (of size up to 10 fm), but the case of compact tetraquark states is not excluded.

\begin{figure*}[hbt]
\centerline{\includegraphics[width=0.58\textwidth]{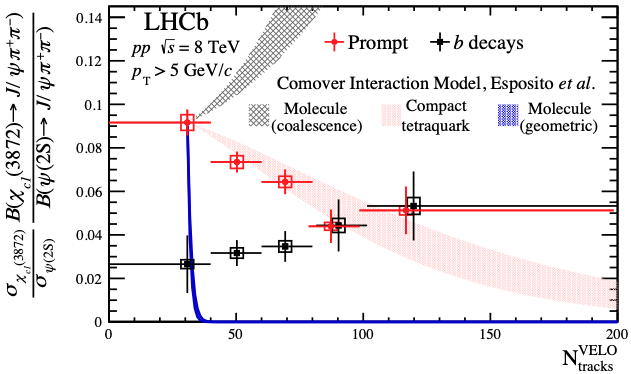} \includegraphics[width=0.42\textwidth]{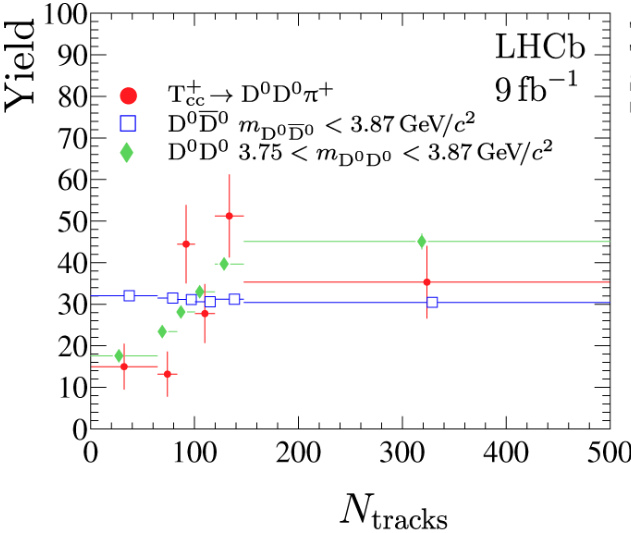}}
\caption{Multiplicity dependence of the cross section ratio of $\chi_{c1}(3872)$ and $\psi$ meson (left) and of the $T_{cc}^+$ measured yield \cite{LHCb:2021auc} (right).}
\label{fig:had4}
\end{figure*}

The study of exotic-hadron production can possibly also shed light on hadronization mechanisms, as suggested by the recent observation by LHCb \cite{LHCb:2020sey} of an event-multiplicity dependence of the $\chi_{c1}(3872)$ meson production in pp collisions, shown in Fig.~\ref{fig:had4} (left). The measurement is described in the comover model~\cite{Esposito:2020ywk} as break-up via interaction with other particles, modeled with an effective cross section dependent on the size of the hadron. A relatively-compact tetraquark is preferred in this case, but it is currently discussed how conclusive this comparison is~\cite{Wu:2020zbx,Lee:2023ysk}. 
 The cross section ratio was observed to be larger in pPb collisions \cite{LHCb-CONF-2022-001}.
The multiplicity dependence of $T_{cc}^+$ production \cite{LHCb:2021auc} suggests an increasing trend (Fig.~\ref{fig:had4} right), which is surprising, given the comparable size of the two hadrons.

Another very interesting exotic state, the $X(6900)$, discovered in 2020 by LHCb \cite{LHCb:2020bwg} and recently confirmed by ATLAS~\cite{ATLAS:2023bft} and CMS~\cite{CMS:2023owd}, is a candidate for a tetraquark of charm quarks and antiquarks. Studying it in the future as a function of multiplicity, challenging as it certainly is, will be an important component towards understanding the hadronization of complex exotic hadrons.

\subsection{Next Steps\label{sec:hadNext}}

In this section the possible next steps discussed at the workshop are presented (essentially a wish list for the experiments, clearly a long-term program):
\begin{itemize}
\item  Other non-strange baryon than protons vs. multiplicity (which points to the $\Delta$ resonances, experimentally a very challenging measurement).
\item Correlations (at low momentum) between baryons, both with light and heavy quarks. 
\item $\Lc/\Dzero$ ratio at low multiplicities (and at high \pt) in pp collisions, to check if the value in \ee is recovered.
\item Clarifying (experimentally) the apparent inconsistency in the $\Lc/\Dzero$ dependence on rapidity.
\item The measurement of $\Lambda_c$ elliptic flow in PbPb and its difference w.r.t. the D one would allow to infer the relevance of diquark degrees of freedom in the QGP, see Fig.~\ref{fig:had3}. 

\item Search for new excited baryon states, both for the charm and bottom sector.
\item More differential $\chi_{c1}(3872)$ measurements, in yield rather than ratio to $\PsiTwos$.
\end{itemize}

\newcommand{\pT} {\ensuremath{p_{\rm T}}}
\newcommand{\jpsi} {\ensuremath{\mathrm{J}/\psi}\xspace}
\newcommand{\RAA} {\ensuremath{R_{\rm AA}}}
\newcommand{\vv} {\ensuremath{v_2}}

\section{Energy loss and transport from small to large systems}
\label{sec:energyloss}
Despite the impressive progresses on both the experimental and theoretical side in the last twenty years, a fully satisfactory description of in-medium energy loss and related phenomena has not been reached yet. Still challenging is the comprehensive description of $\RAA$ and flow data of several particles, which provide sensitivity to the energy-loss process and to its dependence on parton colour-charge and mass. Furthermore, the observation of collective effects in small systems without sizeable signs of energy loss is a major puzzle yet to be solved. 

\subsection{Signals for energy loss in small systems}
One of the most exciting challenges posed by the experimental results obtained in 
small collision systems is the so-called $\vv-\RpPb$ 
puzzle. Finite elliptic flow, $\vv > 0$, is  observed for heavy flavour 
hadrons~\cite{CMS:2020qul, ATLAS:2019xqc} and even for 
jets~\cite{ALICE:2022cwa,ATLAS:2023bmp}, which could be naturally explained  by 
partonic energy loss, while \RpPb is compatible with nuclear PDF model 
calculations~\cite{ALICE:2021wct,CMS:2016svx,ATLAS:2014cpa,LHCb:2022rlh}. 
While such a behavior is predicted by the CGC model, it also predicts bottonium $\vv (\Upsilon ) > 0$~\cite{Zhang:2019dth}, which is not observed~\cite{CMS:2022gdo}. 
Other possible solutions to the $\vv-\RpPb$ puzzle, which do not include energy 
loss, exist. For instance, AMPT predicts  $\vv  > 0$ and $\RpPb \approx 1$ by means 
of a parton escape mechanism (see e.g.~\cite{ALICE:2022cwa}). Moreover, a Glasma 
phase alone could give “diffusion” and no energy loss resulting in $\RpPb > 1$ which 
moves back to unity through energy loss in a medium \cite{Ipp:2020mjc,Liu_2021}.

Taking into account experimental uncertainties, \RpPb\ provides only relatively weak 
constraints on possible energy loss effects, and thus they cannot  be completely excluded. Employing 
jet-hadron correlation measurements in pPb collisions for 
charged jet energies in the range 
$15-60$~GeV/$c$, ALICE has provided a  limit on the energy 
lost outside a jet cone of $R = 0.4$ 
of 0.4~GeV at 90\% CL, ATLAS has extended the measurements to jet $\pT > 60$~GeV/$c$ 
without observing a clear energy loss pattern. With the larger data samples 
projected  to be collected during the LHC Run 3 for 
pp, pPb and OO collisions this limit can be pushed below 0.1~GeV~\cite{ALICE-PUBLIC-2021-004}.

Energy loss effects are expected to be larger in collisions producing a large 
particle multiplicity, which is interpreted as due to the high number of 
initial parton-parton interactions, and thus, high multiplicity events 
correspond to high initial energy densities. However, the analysis of such 
events is complicated due to selection biases \cite{ALICE:2022qxg}.  An example 
is the jet-hadron azimuth angular correlation measurements in pp, where the 
observed broadening can be reproduced by PYTHIA8  calculations which do not 
include any medium effects~\cite{Krizek:2023sze}.

New ideas for the analysis of pp collisions discussed during the workshop are a better control 
of the multiplicity dispersion over the rapidity covered by the signal and multiplicity measurements regions and search for signals of 
redistributed energy in the underlying event of jets. From the theory side, it would be important that 
models predicting a finite $v_2 > 0$ in small systems provide also the minimum parton energy 
loss that could explain the effect.

\subsection{Role of the pre-equilibrium stage}
Most model calculations of parton energy loss in heavy-ion collisions assume no  interactions in the first $\sim 1$~fm/$c$, i.e. before the QGP formation. However, recent studies in the CGC and in effective kinetic theory predict a  substantial $\hat{q}$ in the pre-hydrodynamics phases \cite{Ipp:2020mjc,Boguslavski:2023alu}, which may have consequences for the determination of $\hat{q}$ in the QGP phase and the understanding of jet quenching (see e.g.~\cite{Andres:2022bql}).

Recent energy loss calculations in heavy-ion collisions consider three simplified scenarios for the initial stages: (1) parton  production time, $\tau_{\rm p}$, and medium production time, $\tau_{\rm m}$ are equal and 0, (2)  $\tau_{\rm p} = \tau_{\rm m} = 1\; {\rm 
fm}/c$ and  (3) $\tau_{\rm p} = 0$ and $\tau_{\rm m} = 1\; {\rm fm}/c$. While all the scenarios can reproduce the $\pT$-dependence of the $R_{\rm PbPb}$, albeit with different coupling parameters, once the coupling parameter is fixed, they yield a significantly different high-\pT\ particle $v_2$. This highlights the importance of the treatment of energy loss in the initial stages to properly understand the $R_{\rm PbPb}$ and high-$p_T$ $v_2$ data.

While a large $\hat q$ is predicted in the region $0-1$~fm/$c$, so far no realistic medium-interactions have been considered in this pre-hydrodynamics phase.  We think that efforts from the theory community are needed to understand the consequences for $v_2$ and energy loss. Moreover, we wonder whether small systems can be considered a  proxy for pre-equilibrium effects since 
in these systems this phase dominates.

\subsection{Energy dependence of $\hat{q}$}
Another interesting theoretical result was brought to our attention during the 
workshop: rigorous analytical estimations within the framework of kinetic 
theory have revealed significant jet energy dependence of $\hat{q}$. For the 
first time, all possible diagrams for $2\rightarrow 
2$~\cite{Grishmanovskii:2022tpb} and $2\rightarrow 3$ ~\cite{Song:2022wil, 
Grishmanovskii:2023gog} quark-gluon scatterings are being thoroughly utilized 
to evaluate $\hat{q}$ within the Dynamical Quasi-Particle Model, which describes 
the QGP phase in the PHSD transport approach \cite{Moreau:2019vhw}. At a high-temperature limit obtained $E_{jet}$-dependence is in accordance with the well-known LO-HTL leading-log expression \cite{Arnold:2008vd, Caron-Huot:2008zna}. 
This energy dependence is sensitive to the choice of the coupling constant 
and it can be especially useful for further improvements of jet-quenching models, which usually do not take $E_{jet}$-dependence into account or fail to reproduce asymptotic leading-log behavior.

We think it is important to quantify the consequences for the in-medium parton shower evolution and jet 
shapes.  Can the energy dependence be constrained from experimental data?

\subsection{Energy loss in Quarkonium production}
The observation that in pp collisions prompt \jpsi\ mesons are part of jets and carry, on 
average, only 50\% of the jet energy came as a surprise, since event generators predict them to be produced almost isolated. A possible 
explanation is that \jpsi\ pre-states are produced 
as colour-octets which radiate gluons evolving into a colour singlet. Also \jpsi mesons produced as part of 
a parton shower with a 
$g \rightarrow c \bar c$ splitting are non-isolated. In both cases, the directly production of 
\jpsi mesons can be modified by the presence of a QGP. Indeed, CMS observes that in PbPb collisions high-\pT\ non-isolated  \jpsi are more suppressed than the isolated ones \cite{CMS:2021puf}.

So far nothing is known about the nature of the particles produced in association with 
the \jpsi. We think it is important to characterise them measuring their \pT\ and angular distributions 
(or studying jet shapes) in order to constrain the production mechanism. One should also search for 
modifications of the jet structure comparing PbPb and pp collisions. Do the same phenomena occur for 
$\Upsilon$ production? Moreover, the mechanisms responsible for high-\pT\ \jpsi\ suppression might also 
play a role in charmed hadron suppression, since the $c \bar c$ pair stays for some time in a colour 
octet state. We propose to  
study charm suppression as a function of the ${\rm D}\bar {\rm D}$ opening angle.

\subsection{Better constraints on Heavy Flavor Diffusion}
In heavy-ion collisions, charm hadron spectra in the low-\pT\ region and $v_2$ are used to constrain 
the charm diffusion coefficient. However, collisional energy loss is not the only key ingredient for 
the calculations. It has been shown that 
without radiative energy loss, the predicted \RAA\ is too high and $v_2$ too low~\cite{ALICE:2021rxa, Song:2022wil}.
Another important ingredient for the interpretation of the experimental results is the relation between 
the $c$-quark and the hadron \pT. This relation is different for hadronisation via recombination and 
fragmentation. For recombination the momentum of the hadron is larger than that of the charm quark and 
it also inherits part of the flow of the light hadron with which it combines \cite{Song:2015sfa}.

We think that in order to mitigate the influence  of the hadronisation mechanism on 
the determination of the diffusion coefficient one should limit the analysis to models which can 
describe simultaneously \RAA, $v_2$ and the baryon-to-meson ratio $\Lambda_c / 
{\rm D}$. Further improvements will be obtained by extending the measurements to 
lower $\pT$. In the future, additional observables such as 
${\rm D}^0$-hadron $v_2$ with event shape engineering (ESE), ${\rm D}^0$ $v_2 \{4\}$, 
and  $v_3$ as well as correlations of ${\rm D}^0$ versus pion $v_n$ in ESE classes 
will improve the constraints on the HF diffusion \cite{Plumari:2019hzp,Sambataro:2022sns}.
Moreover, we emphasise the importance for models to provide 
uncertainty bands since they have a large effect on $\chi^2/ndf$.

Ultimately, one will use B mesons (LHC Run 3 and beyond, sPHENIX) to measure the 
heavy-flavour diffusion coefficient. For $b$-quarks, there are less uncertainties 
in the transport description (Boltzmann, Langevin ..) \cite{Prino_2016, zhao2020heavy}.
Furthermore, the larger mass brings the calculations closer to the infinite-mass 
approximation used in lattice QCD calculations of the diffusion coefficient \cite{Sambataro:2023tlv}.

\subsection{Signals for Heavy Flavor Thermalisation}
Modifications of charm hadron spectra and elliptic flow 
represent only indirect signs of 
charm thermalisation. Thermalisation implies that the 
information about the 
initial conditions of the production gets blurred 
completely. Ultimately, this can 
be demonstrated with $\rm D\overline D$ azimuthal angle correlations. The 
relative height of the 
nearside and back-to-back peaks is sensitive to energy 
loss and the 
width of these peaks to the extent of thermalisation, which would imply full isotropisation at low $\pT$ (see e.g.~\cite{Nahrgang:2013saa}).

These measurements are very statistics hungry and might be only possible with the 
ALICE\,3 detector in LHC Run 5~\cite{ALICE:2022wwr}. It would be interesting to see how much one can 
already learn from charm hadron--inclusive hadron correlations. On the one hand, the 
decorrelation from the decay renders the measurement less sensitive, on the other 
hand one gains a large factor in sample size. Moreover, experimental performance 
studies need calculations with state-of-the-art models.

\subsection{Challenges for future high precision measurements }
We close this chapter with a section on more general considerations concerning 
the challenges for future high precision measurements.
Today one witnesses a significant variability among models with different description of partonic interactions in the medium, medium evolution, hadronisation, hadronic phase, and nuclear PDFs. Important questions to be answered by 
theorists are: whether the choice and number of modelling parameters are under 
sufficient control, and whether the observables currently provided by the experiments are already 
optimal.

Results from Bayesian analyses for parameter constraints are crucially sensitive to uncertainties:
wrong uncertainties equals to wrong results. Hence, experimentalists must assure that the uncertainties they estimate are under sufficient control for the high-precision era. For instance, the tendency to choose systematic uncertainties conservatively can lead to overestimated overall uncertainties 
entering the fits. A more complicated issue are correlations among systematic 
uncertainties, since in many cases they are 
not evaluated within the same measured distribution and very rarely among different measurements 
performed by different groups. Last but not least,  models are sometimes used to evaluate the systematic 
uncertainties, which also introduces correlations between the experimental uncertainties and the models.

\section{QCD and astrophysics}
\label{sec:astrophysics}
\newcommand{\lna}{\ln\! A}
\newcommand{\xmax}{X_\text{max}}
\newcommand{\nmu}{N_\mu}
\newcommand{\mlna}{\langle \lna \rangle}
\newcommand{\piz}{\pi^0}
\newcommand{\pO}{pO\xspace}
\newcommand{\Hep}{{He}p\xspace}
\newcommand{\pHe}{p{He}\xspace}
\newcommand{\HeHe}{{He}{He}\xspace}
\newcommand{\sqrtsnn}{\sqrt{s_\text{NN}}}
\newcommand{\deuteron}{d\xspace}
\newcommand{\antiproton}{$\mathrm{\bar{p}}$\xspace}
\newcommand{\antideuteron}{$\mathrm{\bar{d}}$\xspace}
\newcommand{\antihelium}{$\mathrm{\overline{He}}$\xspace}

\begin{table*}
\centering
  \label{tab:astro}
   \caption{\centering Desired input for astroparticle and astrophysics from colliders.}
  \begin{tabular}{p{4cm} p{2.5cm} p{3.5cm}}
    \toprule
    Motivation & Experiments & Desirables \\
    \midrule
    \multicolumn{3}{c}{\it Anti-hyperon puzzle in neutron stars}
    \\
    Interactions of hyperons (Y) and nucleons (N); N$\Sigma$, $\Lambda\phi$; three-body interactions NNY, NYY, YYY
                 &
    ALICE, LHCb
                 &
    Precision measurements, rapidity dependence?
    \\
    \midrule
    \multicolumn{3}{c}{\it Searches for dark matter and primordial antimatter}
    \\
    In p$\mathrm{H}_2$, pHe, p$\mathrm{D}_2$: production of $\bar{d}/\overline{\mathrm{He}}$, $\bar{p}$, $e^+$, $\pi^0 \to \gamma\gamma$, $\eta \to \gamma\gamma$
                 &
    LHCb SMOG2, AMBER, NA61
                 &
    LHC run at 450 GeV
    \\
    \midrule
    \multicolumn{3}{c}{\it Muon puzzle in air showers}
    \\
    Charged particle spectra; pO and $\pi$O inelastic cross-section; multiplicity-dependent hadrochemistry: $R = E(\pi^0)/E_\mathrm{tot}$, $p/\pi$, $\Lambda/K_S^0$, $\Xi/\Lambda$, \dots
                 &
    ALICE, LHCb, LHCf, FOCAL
                 &
    Precision measurements over wide range in $\eta$ and $\sqrt{s}$ in p-ion systems, especially pO
    \\
    \midrule
    \multicolumn{3}{c}{\it Atmospheric background to astro-neutrinos}
    \\
    $\Dzero$, $\Dplus$ production cross-sections at forward rapidity;
    multiplicity-dependent charm production; oxygen gluon nPDF  at $x \sim 10^{-7}$
                 &
    LHCb, FASER
                 &
    Forward acceptance at large rapidity required to reach small $x$, LHC run at 450 GeV
    \\
    \bottomrule
  \end{tabular}
\end{table*}

Along studies of QCD that cast light on the structure and behavior of nucleons and nuclei, various connections with astroparticle and astrophysics-related ``puzzles'' have been discussed during the workshop. A better understanding of phenomena in soft QCD is expected to lead to breakthroughs in these fields. We summarize the input needed from collider experiments for the applications in astroparticle and astrophysics in Table~\ref{tab:astro}.

\subsection{Hyperon puzzle for neutron stars} 
\label{ch:neutronstars}
The interaction of hyperons (Y) with nucleons (N) is one of the keys to understand the composition of the most dense objects in our Universe: neutron stars (NS)~\cite{Ozel:2016oaf,Riley:2019yda}.
These are  characterised by large masses ($M\approx 1.2$--2.2 solar masses $M_\odot$) and small radii ($R\approx 9$--13~km)~\cite{Demorest:2010bx,Antoniadis:2013pzd,Cromartie:2019kug}.
In the standard scenario, the gravitational pressure is counter-balanced by the Fermi pressure of neutrons in the core, which, along with electrons, are the only particles from the parent star remnants.

The high-density environment ($\rho\approx4\,\rho_0$, with $\rho_0$ being the nuclear density) assumed to occur in the interior of NS leads to an increase in the Fermi energy of the nucleons, resulting into the appearance of new degrees of freedom such as hyperons. Nonetheless, this energetically-favoured production of strange hadrons induces a softening of the Equation of State (EoS) incompatible with the current highest mass limit from experimental observations of close to $3\,M_\odot$~\cite{TheLIGOScientific:2017qsa,Lattimer:2019eez,Lattimer:2021emm}.
For this reason, the presence of hyperons inside the inner cores of NS is still under debate, and this ``hyperon puzzle'' is far from being solved~\cite{Djapo:2008au,Tolos:2020aln}.
A possible way out is represented by two-body and three-body repulsive YN and YNN interactions. In both cases, a sufficiently strong YN or YNN repulsive interaction can push the appearance of hyperons to larger densities, limiting the possible presence of these particle species inside NS, stiffening the EoS and leading to larger star masses.

Currently, ALICE femtoscopic measurements in small systems (\pp, pPb) for baryon-baryon pairs involving hyperons deliver the most precise data on the residual strong interaction between nucleons and strange hadrons~\cite{ALICE:2018ysd, ALICE:2019hdt,ALICE:2020mfd, ALICE:2020ibs,  ALICE:2019buq,ALICE:2019eol, ALICE:2021njx, ALICE:2022enj, ALICE:2022uso, ALICE:2022vzr, ALICE:2019gcn, ALICE:2021szj, ALICE:2022yyh}.

The results obtained in recent years from femtoscopic measurements in small colliding systems have proven that femtoscopy can play a central role in understanding the dynamics between hyperons and nucleons in vacuum. For example,  the ALICE measurements on p$\Xi^-$ pairs~\cite{ALICE:2019hdt,ALICE:2020mfd} confirmed a strong attractive interaction between these two hadrons and provided a direct confirmation of lattice potentials~\cite{Sasaki:2019qnh}, leading to a better understanding of the interaction among hyperons and nucleons.

A comparison between hadronic models and these data is necessary in order to constrain calculations at finite density and to pin down the behavior of hyperons in a dense matter environment.
The great possibility to investigate, within the femtoscopy technique, different Y--N interactions and to extend the measurements to three-body forces, can finally provide quantitative input to the long-standing hyperon puzzle.

\subsection{Muon Puzzle in atmospheric showers} 

\begin{figure}
  \centering
  \caption{Compilation of muon measurements converted to the abstract $z$-scale for the hadronic interaction model EPOS-LHC. The $z$-scale is given by $z = (\ln(N_\mu) - \ln(N_{\mu,p})) / ( \ln(N_\mu,\text{Fe}) - \ln(N_{\mu,p}))$, where $N_\mu$ is the measured muon content and $N_{\mu,p}$ and $N_{\mu,\text{Fe}}$ are predictions for proton and iron CRs, respectively. The energy scales of the experimental datasets shown here have been cross-calibrated as described in Ref.\,\cite{Soldin:2021wyv}. The expected value $z_\text{mass}$ based on the cosmic-ray mass composition estimated by the GSF model \cite{Dembinski:2017zsh} is subtracted. Also shown are $z_\text{mass}$ values computed from $\xmax$ measurements by the Pierre Auger Observatory (grey band).}
  \label{fig:z_rescaled}
  \includegraphics[width=0.5\textwidth]{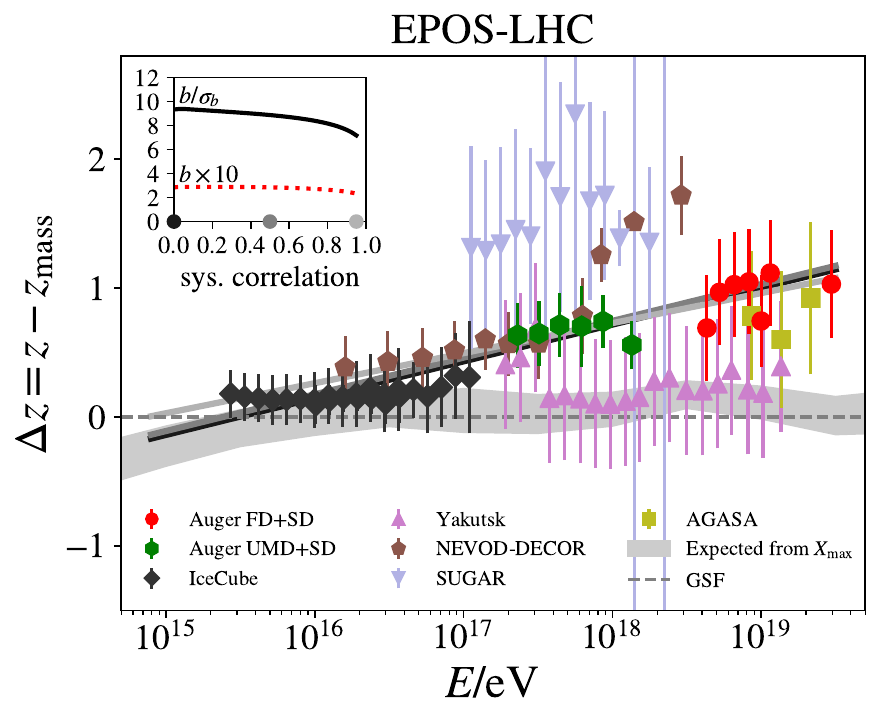}
\end{figure}

Cosmic rays (CRs) are fully ionized nuclei that arrive at Earth with relativistic kinetic energies. At energies above 100 TeV, CRs are observed indirectly by ground-based experiments through extensive air showers, particle showers produced in Earth's atmosphere.
Air shower features are used to determine the direction, energy, and mass of the cosmic ray. Two main features of an air shower are used to estimate the mass, its depth of shower maximum $\xmax$, and the number of muons $\nmu$ produced in the shower. The bulk of muons is produced at the end of a hadronic cascade dominated by interactions in the non-perturbative regime of QCD, which are modeled by effective theories like the core-corona model EPOS-LHC \cite{Pierog:2009zt,Pierog:2013ria},
and the parton shower model Pythia with string-string interactions \cite{Bierlich:2014xba,Bierlich:2015rha,Bierlich:2017vhg}. Especially thanks to the latest generation of large air shower experiments that measure air showers in unprecedented detail, a significant muon deficit is observed in current state-of-the-art simulations based on $\xmax$ in comparison to observations in real air showers (Fig.\,\ref{fig:z_rescaled}). This muon production anomaly is called the ``Muon Puzzle in air showers'', more information can be found in Ref.\,\cite{Albrecht:2021cxw}. The produced number of muons is very sensitive to the energy fraction carried away by photons, which are primarily produced in $\piz$ decays \cite{Ulrich:2010rg,Baur:2019cpv,Pierog:2021sql}. Small changes of the $\piz$-fraction among all produced hadrons in the forward region have a large effect on muon production \cite{Baur:2019cpv,Anchordoqui:2019laz}.

To solve the Muon Puzzle, forward measurements of the light-flavor hadron production at small transverse momentum in minimum bias events are needed, in particular of hadrochemistry. Indirectly, precision measurements of the inelastic cross-section in pA collisions are also needed, and ideally also the proton-$\pi$ inelastic cross-section, which affect predictions for $\xmax$. Particularly important are measurements by forward experiments like LHCf \cite{LHCf:2015rcj,LHCf:2017fnw,LHCf:2020hjf}, CMS with CASTOR \cite{CMS:2019kap}, and LHCb \cite{LHCb:2021abm,LHCb:2021vww,LHCb:2022vfn}. A key ingredient to resolve the muon puzzle could be multiplicity-dependent strangeness enhancement observed by ALICE \cite{ALICE:2016fzo} in the central collision region. LHCb found evidence for this effect also in the forward region \cite{LHCb:2022syj}. In the near future, the study of soft hadron production in \pO collisions at the LHC is another key step forward, which provides the best reference for cosmic proton with air, and will greatly reduce the uncertainty coming from interpolations of \pp and pPb data.

\subsection{Production of secondary particles in the Galaxy} 

With the last generation of particle detectors in space (PAMELA~\cite{Adriani_2010}, AMS-02~\cite{PhysRevLett.122.041102}, DAMPE~\cite{DAMPE_2017}, CALET~\cite{Aguilar:2016kjl}, Fermi-LAT~\cite{Fermi-LAT:2009ihh}) and with the future ones like GAPS~\cite{Aramaki_2016}, physics of CRs and gamma rays ($\gamma$ rays) is more and more a precision discipline. The fluxes of positrons ($\mathrm{e}^{+}$)~\cite{PhysRevLett.122.041102} and light antinuclei like antiprotons (\antiproton)~\cite{Aguilar:2016kjl}, and potentially in the near future antideuteron (\antideuteron) and antihelium (\antihelium), are being measured with percent uncertainties in a wide energy range and compared to the expectations from secondary-only production, namely  primary CR interactions with the interstellar medium (ISM), to evidence primary astrophysical sources or contributions from dark matter annihilation or decay~\cite{Giesen_2015, Cuoco_2019, PhysRevD.103.123005, Orusa_2021}. To improve the secondary production description, and infer conclusions on the primary contributions, the uncertainties on hadronic cross-section, currently contributing as $\pm20$\% \cite{Korsmeier_2018} to the secondary \antiproton flux, need to scale down to the  experimental precision. Cross-section measurements in the \pp channel, dominating the secondary  production~\cite{Korsmeier_2018,Orusa_2022,orusa2023new}, or either with the CR projectile or the ISM target replaced by helium (\Hep, \pHe, and \HeHe), are hence needed. Both direct and indirect production have to be addressed. For \antiproton, with the dominant contributions coming from prompt emission, hyperon-induced and the possible isospin asymmetry for antineutron production have to be constrained on data~\cite{Winkler_2017}. For \antideuteron and \antihelium nuclei, suppressed in the CRs-ISM production at low energies with respect to DM signal predictions\cite{Donato_2000, Cirelli_2014}, in addition to direct production via coalescence, a possible contribution from ${\overline{\mathrm{\ensuremath{\Lambda}}}}_{b}$ baryon decays~\cite{Winkler_Lb} has to be constrained. For $\mathrm{e}^{+}$ above 1 GeV energy, the flux data by AMS-02 are higher than the current secondary-only predictions ~\cite{dimauro2023novel}, that could be improved with  measurements of the
$\pi^{\pm}$ and $K^{\pm}$ spallation cross-sections from \Hep, \pHe, and \HeHe collisions.

Most of the $\gamma$ rays produced by hadronic interactions detected by Fermi-LAT~\cite{Fermi-LAT:2022byn} originate instead from the $\pi^{0} \rightarrow \gamma \gamma$ decay and new data on the Lorentz invariant cross section for $\pi^0$ production are needed~\cite{orusa2023new} to bring down the $\sigma(pp \rightarrow \pi^0 \,X)$ uncertainty to the same level as the Fermi-LAT statistical errors. Particularly, measurements for $p_\text{T} \lesssim$ 1 GeV, large coverage in $x_\text{F}=2 p_\text{L}/\sqrt{s}$ and beam energies in the laboratory frame from a few tens of GeV to few TeV, both for \pp and \pHe collisions, would contribute most in decreasing the uncertainties.

In parallel to the theoretical progresses, experiments from different facilities are providing relevant measurements. At the SPS accelerator at CERN, NA61~\cite{Aduszkiewicz:2017sei} and COMPASS~\cite{adams2019letter} fixed-target experiments are being upgraded to extend and improve \antiproton, $\mathrm{e}^{+}$, and $\gamma$ production cross-section measurements from \pp and \pHe interactions. Leveraging on the injection of gases in the LHC beam-pipe with a system called SMOG, the LHCb experiment has also been operated in fixed-target mode since 2015, providing the first measurements for both direct~\cite{LHCb-PAPER-2018-031} and hyperon-induced~\cite{LHCb-PAPER-2022-006} $\sigma(\mathrm{pHe} \rightarrow \mathrm{\bar{p}}\,X, \sqrtsnn = 110\,GeV)$. Measurements of \antideuteron, $K^\pm$, $\pi^0$, $\eta$, $\mathrm{e}^{+}$ in the same dataset and of \antiproton at lower beam energies are also ongoing or feasible with the SMOG system. With the upgrade of the LHCb fixed-target programme, namely the installation of a gas storage cell upstream of the LHCb nominal interaction point and a new gas feed system~\cite{LHCb-TDR-020}, hydrogen and deuterium could be injected as well, shedding light on the possible isospin asymmetry in the antineutron-to-antiproton production. Moreover, concurrent data-taking with the beam-beam collisions was demonstrated to be feasible~\cite{CERN-THESIS-2021-313}, resulting in larger collected data samples and more precise cross-section measurements~\cite{LHCb-PUB-2018-015}. All of this is resulting in a unique facility for QCD measurements at the LHC and, notably, for its connection to astrophysics.

\section{Summary and outlook}
\label{sec:summary}
The $4^{\mathrm{th}}$ edition of the workshop QCD challenges from pp to AA collisions took place in Padua, Italy, from $13^{\mathrm{th}}$ to $17^{\mathrm{th}}$ February 2023. It was attended by 65 researchers, including both ``senior'' experimental and theoretical physicists and young researchers (about 15\%). The agenda consisted of plenary sessions alternated to parallel round-table sessions organised in parallel on six tracks, with a few sessions dedicated to inter-track discussions. The workshop format was very effective in stimulating discussions. In the round-table sessions there was ample time for following up the topics presented during plenary sessions, the questions raised, as well as for deepening items and ideas that session conveners had prepared in advance. 

In general, the necessity emerging in the last years of shifting away from the paradigm considering proton--proton and heavy-ion collisions as systems featuring different and unrelated properties was further established during the workshop. Understanding the observation of collective flow behavior as well as the dependence of strangeness production on the event charged-particle multiplicity is considered of pivotal importance. It was noted in particular the need for phenomenological approaches that take into account non-equilibrium effects that become more and more important going from large to small collision systems. The performance of core-corona approaches in describing the hadrochemistry and kinematics across collision systems was discussed, together with the possible relevant role played by the core in pp collisions. 
Gaining a realistic three-dimensional description of the entire interaction area of the collision was highlighted as a necessary step for interpreting rapidity-differential measurements and observables like $v_{1}$ and particle asymmetries sensitive to effects and correlations spanning over wide rapidity intervals. Full 3D simulations are also crucial for the comparison to the elliptic and triangular flow data in small systems. Moreover, studying the correlation between fluctuations in mean \pt and elliptic-flow magnitude could help resolving initial-state effects like initial momentum correlations in small systems. 

So far no evidence was found of energy loss in small systems while positive elliptic-flow values have been observed for several high-energy probes, like heavy-flavour, quarkonia and jets. Also with jet-hadron correlations, expected to be less affected by the experimental uncertainties, only upper limits to the out-of-cone radiation could be set in pPb collisions. A better control of the multiplicity dispersion over large rapidity regions and searches for signals of the redistributed energy in the underlying event were discussed as possible directions to solve the $\RpPb-v_{2}$ puzzle. On the theory side, it would be important that models predicting positive $v_{2}$ specified also the minimum parton energy loss required to generate it. Sensitivity to possible energy loss in the final state requires understanding of the initial state also for the production of high-energy partons. At high collision energy the colliding nuclei can be represented as dense gluonic systems. Saturation effects, which can be described by the Colour Glass Condensate effective field theory, can become relevant when low Bjorken-$x$ values are involved. As reviewed during the workshop, in recent years, substantial progress has been made to constrain this low-$x$ regime with measurements of particle production at forward rapidity in pA collisions and of photoproduction of vector mesons in ultraperipheral collisions. These data stimulated significant theoretical developments, especially to overcome the obstacles posed by the poor stability of perturbative calculations of the amplitude to the inclusion of heavy-vector meson data in global PDF analyses. However, attempts of extracting proton PDF by relating PDF to GPD and using HERA and LHCb \Jpsi photoproduction data show large differences from PDF determined using LHCb $\Dzero$-meson data. This discrepancy indicates the need of additional theoretical work but also that new experimental data including those from pO and OO runs at the LHC and forward direct-photon production measurements that detector upgrades will allow, will be vital to observe and understand nonlinear QCD effects.

Similarly to the appearance of long-range correlations and flow-like effects, the hadronisation process is a further important case in which concepts originally introduced to describe hadron production yields and momentum distributions in heavy-ion collisions turned out to be effective for proton--proton and proton--Pb collisions. In particular, models including quark-recombination as a hadronisation process or based on a statistical approach agnostic of any partonic phase describe heavy-flavour baryon production significantly better than calculations that employ fragmentation functions constrained to reproduce $\ee$ data and that strongly underestimate the baryon-to-meson ratios. Models based on string-fragmentation expect an enhancement of baryon-to-meson ratios in hadronic collisions only if colour reconnection is implemented more realistically, going beyond the leading-colour approximation that is sufficient for describe \ee data. It is expected that the experiments will make an effort in the incoming years to measure the production yields of $\SigmacZeroPlus$ and of the charm-and-strange baryons $\XicPlusZero$ and $\Omegac$ with improved precision over a wider transverse momentum range and better profiling their evolution with multiplicity across collision systems. A precise measurement of beauty baryon production at midrapidity, which is currently missing, will allow to determine whether baryons are effectively less produced at forward rapidity, resolving the apparent tension between ALICE and LHCb charm baryon data. 
On the theoretical side, a careful study of the expected multiplicity dependence in small systems as well as a study of the possible rapidity dependence of the baryon-to-meson ratios is desirable. Particle angular correlations are also envisaged as powerful probes of the hadronisation dynamics. Aside ``standard'' heavy-flavour baryons, studying the production yields and the properties of exotic hadrons with heavy-quark content may add further insight into hadronisation, as well as into the nature and internal structures of these states.

For \enquote{large} collision systems and for the characterisation of the QGP, polarisation measurements are seen as a very promising way to investigate vorticity phenomena that can arise in the medium both as a global effect related to the medium rotation as well as localised ones in response to the passage of a high-energy particle. The possibility to connect quark-spin alignment to hadron polarisation is based on the assumption that a sizeable fraction of hadrons is formed via coalescence, inheriting the spin of the system of recombining quarks. Polarization measurements can therefore also probe the hadronisation process. Constraining the hadronisation is also fundamental for making a step further in the study of in-medium energy loss and heavy-quark transport. In particular, for determining the spatial diffusion coefficient from charm-hadron data, it is mandatory to rely only on models that can simultaneously describe $R_{\rm AA}$, $v_{2}$ and the $\LcD$ ratio. Additional constraints could be set by new and precise measurements of D-meson flow coefficients with Event Shape Engineering or of direct correlations of D-meson and pion flow coefficients. Eventually, thanks to the detector upgrades and the expected new measurements of beauty flow and $\Raa$, the spatial-diffusion coefficient could be determined with beauty hadrons. This possibility has the advantage that the uncertainties in both transport models and lattice-QCD calculations are smaller on beauty observables than on charm ones. 

Concerning in-medium energy loss, the energy dependence of the $\hat{q}$ coefficient, which encodes the scattering power of the medium, was discussed in the workshop. Understanding its consequences for in-medium parton shower and jet shapes and how experimental data could constrain it are important challenges for the future. It was also highlighted that treating energy loss in the first instants after the collision, within the pre-equilibrium stage of the QGP, an interval often neglected in energy-loss calculations, could be important for the simultaneous description of $R_{\rm AA}$ and $v_{2}$ and for the determination of $\hat{q}$. It was also proposed to investigate whether small systems could provide a proxy for addressing the pre-equilibrium stage.  
As a different direction to study energy loss and in particular medium-induced radiation, the potential of jet substructure measurements for determining the characteristic angle of medium-induced emissions, $\theta_{c}$ was extensively debated in the workshop. Observables based on clustering-tree, like the dynamically groomed jet radius, and on energy-flow, like energy-energy correlators, were discussed. For the latter, measurements in $\gamma$-jet events were advertised as promising thanks to mitigation of selection-bias effects. In general, isolating the $\theta_{c}$-related dynamics remains a very challenging task: jet-substructure observables in high-\pt jets could more naturally match the requirements identified as necessary to fulfill this task, i.e. having a calculable vacuum baseline, resilience to the underlying event, and reduced sensitivity to medium response.

At the workshop some QCD aspects influencing the understanding of astrophysics and cosmic-ray physics were also discussed. Far from being solved is the "hyperon puzzle" for neutron stars. It was remarked that there is a need of further femtoscopic measurements for clarifying whether two-body or three-body repulsive interactions exist and limit the formation of hyperons in the core of neutron stars, stiffening the equation of state and allowing larger star masses. Another "puzzle" discussed in the workshop concerns the deficit of muons in state-of-the-art simulations compared to observations in air showers. Important inputs from particle-accelerator experiments are forward measurements of light-flavor hadron production at small transverse momentum, in particular studies of hadrochemistry, including profiling strangeness enhancement as a function of the event multiplicity. It was noted that a future run of pO collisions at the LHC could provide a better reference for cosmic rays than current pp and pPb data. Concerning the search of dark-matter signals, the constantly improving precision of measurements of charged cosmic rays and $\gamma$ rays from particle detectors in space is opening a new era. However, in order to draw conclusions on the primary sources which the description of the secondary production in cosmic-ray interactions with the interstellar medium must be improved. To this purpose, precise measurements of the production cross section of $\overline{\rm p}$, $\overline{\rm d}$, $\overline{\rm He}$, $\mathrm{K}^{\pm}$, $\pi^{0}$, $\mathrm{\eta}$ in pp as well as in pHe, Hep, and HeHe collisions possible with the NA61 upgrade, the COMPASS++/AMBER experiment, and the LHCb SMOG program, are fundamental to reduce the uncertainties on hadronic cross sections. Constraining light antinuclei production in beauty-hadron decay with LHCb or in the future with ALICE3 will also be important.

We would like to conclude this document with few final remarks on the effectiveness of the workshop structure. With respect to previous editions, an effort was made to enlarge the participation of physicists working on QCD-related topics outside the heavy-ion physics community. This was a major objective of this edition that we think should be pursued also in the future. The introduction of a session connecting QCD at colliders with astrophysics themes was also a new ingredient of this edition appreciated by the participants. For young researchers, the participation in the round-table sessions was a particularly valuable experience. During the organisation of the event, it was deliberately chosen to give session conveners ample freedom to define the topics they preferred to deepen during the sessions and, to a large extent, define the speakers. While this freedom assures lively and focused discussions, it also makes it more complex to assure some continuity between the topics discussed in different editions. However, we do not consider this as a significant limitation, rather as a feature characterising what should be the expectations and the outcome of a similar event, which, on top of stimulating detailed discussions on well-established open points in the field, is also meant to start new collaborations and directions of research. The organisation of a 1-day online ``prequel'' event a couple of months before the workshop could be useful to make new participants familiar with the workshop format and to sharpen the discussion topics, also defining points and questions to be addressed in the workshop.

\section*{Acknowledgments}
We acknowledge the financial support of the Italian Institute for Nuclear Physics (INFN) and the University of Padua.
This project has received funding from the European Union's Horizon 2020 research and innovation programme under grant agreement N. 824093 (STRONG 2020 project).
We acknowledge the support from the project “CLASH: Pinning down the origin of collective effects in small collision systems”.
The work was further supported by the Czech Science Foundation (GA ČR), project No. 22-27262S. We acknowledge also the support of the Ministry of Education and Science and National Science Centre (Poland) and the U.S. National Science Foundation. L.O. acknowledges the Next Generation action of the European Commission and the MUR funding (PNRR Missione 4) under the HEFESTUS project. P.P. acknowledges financial support from the Academy of Finland project 330448, the Center of Excellence in Quark Matter of the Academy of Finland and the European Research Council project ERC-2018-ADG-835105 YoctoLHC. C.A. has received funding from the European Union’s Horizon 2020 research and innovation program under the Marie Sklodowska-Curie grant agreement No 893021 (JQ4LHC). Finally we acknowledge the grant ERC-CZ LL2327. 

\bibliography{sn-bibliography}

\end{document}